\documentclass[11pt]{article}
\pdfoutput=1
\usepackage{amsmath,amssymb,amsthm}
\usepackage[backref=page]{hyperref}
\usepackage[smalltableaux,centertableaux]{ytableau}
\usepackage{bbold}
\usepackage{authblk}
\usepackage{color}
\usepackage{graphicx}

\numberwithin{equation}{section}

\newcommand{\id}{\mathbb{1}}
\newcommand{\RR}{\mathbb{R}}
\newcommand{\CC}{\mathbb{C}}
\newcommand{\ZZ}{\mathbb{Z}}
\newcommand{\NN}{\mathbb{N}}
\renewcommand{\Re}{\operatorname{Re}}
\renewcommand{\Im}{\operatorname{Im}}

\DeclareMathOperator{\sgn}{sgn}
\DeclareMathOperator{\Tr}{Tr}
\newcommand{\kernel}{\mathbb{S}}
\newcommand{\pf}{\mathcal{Z}}
\newcommand{\op}{\mathcal{O}}
\renewcommand{\sl}{\mathfrak{sl}(2)}

\newcommand{\pv}{\mathrm{p.\!v.\!}} 

\title{Quantum corrections to the BTZ black hole extremality bound from the conformal bootstrap}
\author{Henry Maxfield\thanks{\href{mailto:hmaxfield@physics.ucsb.edu}\texttt{hmaxfield@physics.ucsb.edu}}}
\affil{Department of Physics, University of California,\\ Santa Barbara, CA 93106, USA}

\begin{document}

\maketitle

\begin{abstract}
	Any unitary compact two-dimensional CFT with $c>1$ and no symmetries beyond Virasoro has a parametrically large density of primary states at large spin for $\bar{h}>\bar{h}_\text{extr}\sim \frac{c-1}{24}$, of a universal form determined by modular invariance. By including the contribution of light primary operators and multi-twist composites constructed from them in the modular bootstrap, we find that $\bar{h}_\text{extr}$ receives corrections in a large spin expansion, which we compute at finite $c$. The analysis uses a formulation of the modular S-transform as a Fourier transform acting on the density of primary states. For theories with gravitational duals, $\bar{h}_\text{extr}$ is interpreted as the extremality bound of rotating BTZ black holes, receiving quantum corrections which we compute at one loop by prohibiting naked singularities in the quantum-corrected geometry. This gravity result is reproduced by modular bootstrap in a semiclassical $c\to\infty$ limit.
\end{abstract}

\tableofcontents

\newpage

\section{Introduction}

The bootstrap program applied to two-dimensional conformal field theory has been remarkably successful for the classification and solution of rational theories \cite{Belavin:1984vu}, but far less progress has been made towards understanding the space of theories more generally. Irrational theories, aside from presumably being the generic case, are of particular interest as duals to quantum gravity in AdS$_3$, and are more similar to CFTs in higher dimensions, for which the bootstrap has enjoyed enormous recent progress \cite{1805.04405}. Exact analytic solution or classification of such theories is almost certainly out of reach (with notable exceptions), so our aims must be more modest. Despite these limitations, the constraints of crossing symmetry and modular invariance impose a great deal of rigidity on the spectrum and couplings, even for irrational theories. A wealth of information about infinitely many primary operators can be deduced -- practically, not just in principle -- from very little data about the light spectrum.

In this paper, we will explore one aspect of this, by combining modular invariance of the partition function with recent results concerning the existence and spectrum of composite operators. The results are most universal, and most interesting, for the region of operator dimensions we call the `near-extremal' spectrum. For theories with AdS$_3$ gravitational duals, this corresponds to rapidly rotating BTZ black holes, close to their extremality bound. This bound marks the edge of a range of energy and spin with a parametrically large density of states as determined by the Bekenstein-Hawking formula for black hole entropy, a part of the spectrum that can be treated as effectively continuous for most purposes. Even in generic theories at small central charge, such a continuum of states emerges at large spin, along with a characteristic `extremality bound' marking its edge. Our main results find spin dependence of this bound, determined by the dimensions of light operators, both for generic theories in a large spin expansion and for holographic theories at large central charge.

\begin{figure}[p]
	\centering
	\includegraphics[width=.9\textwidth]{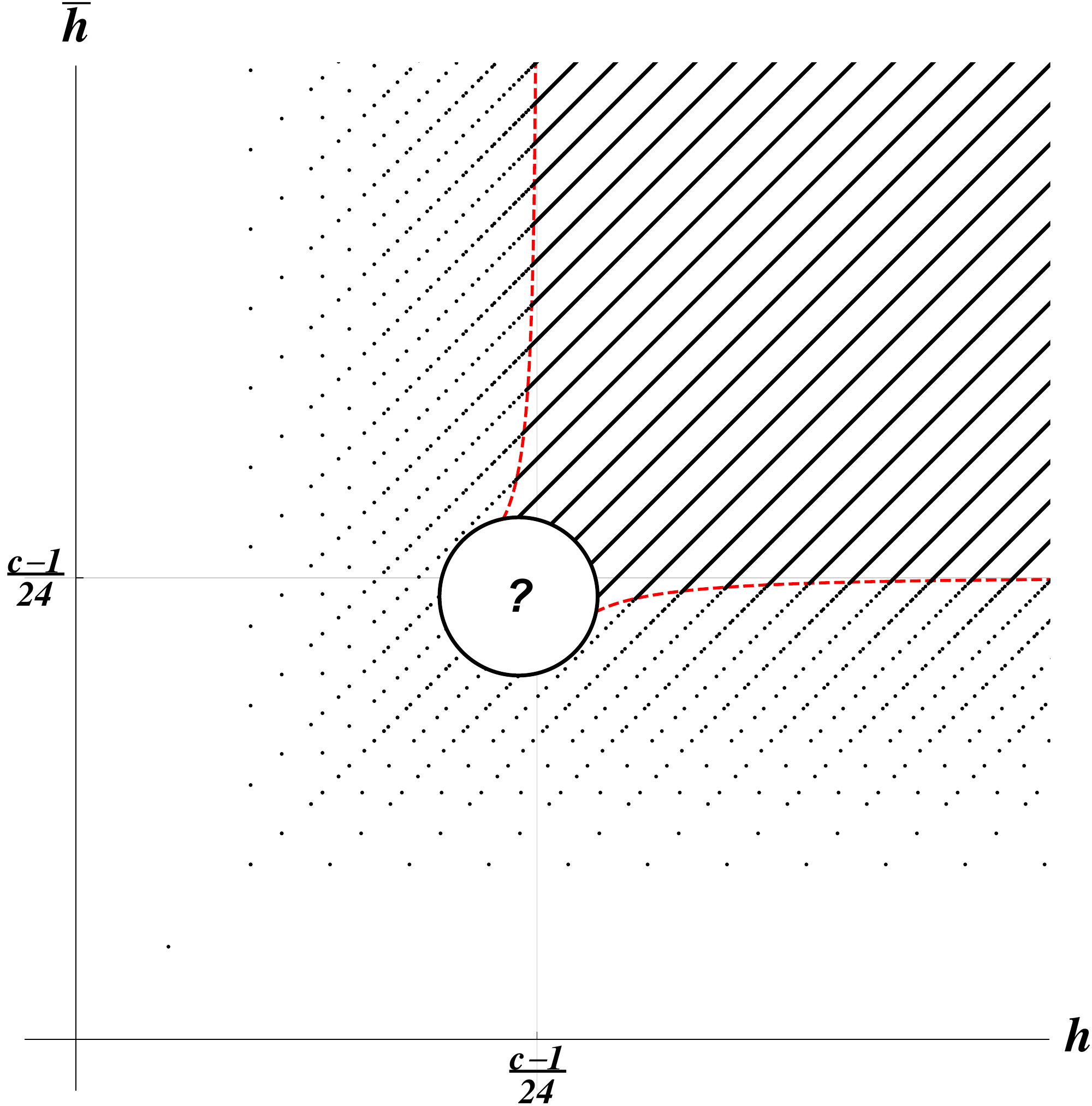}
	\caption{A conjectural cartoon of the spectrum of primaries of an irrational CFT. This paper addresses two features of the spectrum, namely multi-twist operators and the extremality bound, and their relation through modular invariance. The points represent composite multi-twist operators, which grow polynomially in number at large spin. The region above the red dashed line -- the extremality bound -- represents the `continuum' of operators, with entropy growing as $\sqrt{c\ell}$ at large spin $\ell$ or central charge $c$. The twist of light operators determines the shape of the extremality bound.	Generically, we can only trust this picture in an asymptotic expansion in spin, so the region \raisebox{.5pt}{\textcircled{\raisebox{-1.5pt}{?}}} without perturbative control is large. In a theory with weakly coupled AdS$_3$ dual, the perturbation theory is instead controlled by large central charge, the validity is extended, and \raisebox{.5pt}{\textcircled{\raisebox{-1.5pt}{?}}} is small compared to $c$. \label{fig:spectrum}}
\end{figure}

From these results, it is natural to propose a picture of the spectrum of an irrational CFT illustrated in figure \ref{fig:spectrum}, consisting of a small number of `fundamental' operators (single trace in holographic theories), multi-twist composites built from them, and an exponentially large density above an extremality bound.\footnote{There may be additional features of the spectrum which look qualitatively different, such as a dual to finite tension strings in AdS.}
The properties of each of these parts of the spectrum are not independent, but bound together through the bootstrap.

\subsection{Universal results for unitary compact CFTs}

The first part of the paper concerns the spectrum of a generic unitary, compact\footnote{A compact theory is defined to have a discrete spectrum, including a ground state invariant under the $\sl\oplus \sl$ global conformal symmetries.} CFT. We build up a picture of the complete spectrum based on minimal information about a few light states, leveraged with the strong constraints of conformal symmetry.

For this, we take a new approach to the analytic modular bootstrap\footnote{Previous work includes \cite{0902.2790,1307.6562,1608.06241,1905.01319}.}, by formulating modular invariance as a condition on the density of states directly, thereafter dispensing with the partition function. Our strategy will be to choose an appropriate collection of known states of the theory, construct a partition function from them, take a modular S-transform, and express the result as a density of states in the new S-transformed channel. In section \ref{sec:Stransform}, we show that, in appropriate variables, this acts simply as a Fourier transform on the density of primary states (following \cite{hep-th/0101152,1407.6008}). With a judicious choice of the states we start with, there will be a regime of operator dimensions in which the resulting transformed density is indicative of the actual CFT spectrum, because corrections from including more states in the input are parametrically suppressed in that regime.
 
The basic result of this approach -- taking the modular S-transform of the vacuum only -- is the venerable Cardy formula \eqref{eq:Cardy1} \cite{Cardy:1986ie}, giving the density of states at large energy. The same argument (with an additional assumption) gives a formula \eqref{eq:Cardy2} for the asymptotic density of states at large spin, in the limit $h\to\infty$ with $\bar{h}$ held fixed. Including a primary operator besides the vacuum leads to a correction which is suppressed in this limit, as long as the operator in question has $h>0$. The required assumption is therefore a twist gap, meaning in particular that the theory contains no conserved currents (operators with $h=0$) beyond those associated with local conformal invariance.

The most striking aspect of the large spin asymptotic formula is the `extremality bound' alluded to above. This is equivalent to the lower bound on the twist gap derived in \cite{1608.06241,1707.07717} by a closely related method. Fixing $\bar{h}>\frac{c-1}{24}$, the formula implies a large microcanonical entropy growing with spin $\ell=h-\bar{h}$ as $S\sim 2\pi \sqrt{\frac{c-1}{6}\ell}$, but for $\bar{h}<\frac{c-1}{24}$ the density of states grows more slowly, if there are any states at all.  If we take $\bar{h}$ in the `near-extremal' regime, lying slightly above the bound, we find that the density of states at large spin has a square-root edge:
 \begin{equation}
	\rho \approx e^{2\pi \sqrt{\frac{c-1}{6}\ell}}\sqrt{\bar{h}-\frac{c-1}{24}} \qquad \left(\ell\to\infty, \quad 0< \bar{h}-\tfrac{c-1}{24} \ll c^{-1}\right)
\end{equation}

In section \ref{sec:Smultitwist}, we discuss the corrections to this large spin Cardy formula arising from a light operator $\op$ in the theory (initially, the operator of smallest positive twist\footnote{Twist is dimension minus spin, $\tau=\Delta-|\ell|=2\min\{h,\bar{h}\}$; we use the term loosely to refer to any equivalent parameter.}). We could simply include the operator $\op$ itself (and descendants) along with the identity before performing the modular transformation on the spectrum, but we can do better. In addition, there is a tower of infinitely many composite operators built from $\op$, discussed in section \ref{sec:multispec}. We infer the existence and properties of these `multi-trace' operators at large spin from recent results on `double-twist' operators, which used a bootstrap analysis of four-point function crossing \cite{1811.05710,1810.01335}. While the precise spectrum of multi-twist operators depends on details of the theory, at large spin the main features are determined universally, by only the central charge and the dimensions of $\op$.

The modular transform of this universal piece of the multi-twist spectrum has its most important effect in the `near-extremal' regime discussed above, of large spin with $\bar{h}$ close to $\frac{c-1}{24}$. The modular transform of this multi-twist spectrum gives an asymptotic expansion for the density of states, which can be summed into a shift of the `extremality bound' $\bar{h}_\text{extr}$ with a particular spin dependence. For large spin $\ell\gg 1$ and $\bar{h}$ close to $\frac{c-1}{24}$, we have
\begin{equation}\label{eq:hextr1}
\begin{gathered}
	\rho \approx e^{2\pi \sqrt{\frac{c-1}{6}\ell}}\sqrt{\bar{h}-\bar{h}_\text{extr}(\ell)}\\
	 \bar{h}_\text{extr}(\ell) \sim \frac{c-1}{24}-\frac{1}{(2\pi)^2}\frac{e^{-4\pi\alpha P}} {1-e^{-4\pi b P}}
\end{gathered}
\end{equation}
The parameters in this formula are defined in equations \eqref{eq:cParams} and \eqref{eq:hParams}: $P$ is roughly $\sqrt{\ell}$, $\alpha$ is determined by the left-moving dimension $h_{\op}$ of the low-twist operator in question, and $b$ is determined by the central charge. We generalise the result to include the contribution of multiple independent light operators, as well as fermions.

\subsection{Theories with AdS$_3$ gravitational duals}

A particularly important class of irrational CFTs are those with a dual description in terms of gravity in asymptotically AdS$_3$ spacetimes. The formulas for the density of states in the previous section have an interpretation as the Bekenstein-Hawking entropy (with loop corrections) of BTZ black holes \cite{hep-th/9204099,gr-qc/9302012}. In particular, the bound $\bar{h}\geq\frac{c-1}{24}$ corresponds to the extremality bound $M\geq J$, for which there is a classical geometry without naked singularities\footnote{The shift $c\rightarrow c-1$ is a one loop correction from the Casimir energy of gravitons.}.

To make contact with our large spin results, the relevant comparison is to account for the effect of light bulk matter fields on the black hole extremality bound. In section \ref{sec:bulk}, we compute the one-loop expectation value of the stress tensor of a scalar field in the Hartle-Hawking state on the BTZ geometry. Treating this as a source for the linearised Einstein equations, the scalar provides a quantum correction to the geometry, and hence to the `cosmic censorship' criterion that singularities must be hidden behind a horizon. The result is a modified extremality bound:
\begin{equation}\label{eq:hextr2}
	\bar{h}_\text{extr}(\ell) \sim \frac{c-1}{24} -\sum_{n=1}^\infty\frac{1}{(2\pi n)^2}\frac{e^{-2\pi n r_+\Delta}}{1-e^{-4\pi n r_+}}
\end{equation}
Here, $\Delta$ is the conformal dimension of the operator dual to the scalar, and $r_+ = \sqrt{\frac{6\ell}{c}}$ is the horizon radius of the extremal black hole in AdS units, taken to be of order one so that $\ell$ is of order $c$.

In section \ref{sec:SCMI}, we derive the same result (generalised to operators with spin) from a bootstrap argument, applying modular invariance to a spectrum with a non-interacting Bose gas of multi-trace operators, in the limit $c\to\infty$ with $\frac{\ell}{c}$ fixed. The Bose gas includes a larger collection of operators than the results \eqref{eq:hextr1} for a generic theory, for which we included only the leading order in large spin, since this piece does not receive corrections from interactions. This difference accounts for the additional terms in the sum over $n$ appearing in \eqref{eq:hextr2}, but these extra terms are correct only in an approximation where the anomalous dimensions of multi-trace operators can be neglected.

\subsection{Organisation of the paper}

In section \ref{sec:Stransform}, we review and expand upon the modular S-transform as a Fourier transform acting on the density of states. In section \ref{sec:Cardy}, we derive the Cardy formula and its large spin version from this point of view. In section \ref{sec:multispec}, we discuss the spectrum of multi-twist operators. We then apply this to the modular bootstrap in section \ref{sec:Smultitwist}, finding the  result \eqref{eq:hextr1}.

The last two sections, concerned with the semiclassical limit, can be read independently. Section \ref{sec:bulk} describes the calculation of one-loop gravitational corrections to the BTZ extremality bound. Finally, we reproduce this result from modular invariance in the semiclassical limit in section \ref{sec:SCMI}.

In appendix \ref{app:maths}, we give some additional technical discussion of the Fourier transform. In appendix \ref{app:counting}, we count multi-particle states. Appendix \ref{app:Wald} reviews the Wald formalism for conserved quantities, relevant for section \ref{sec:bulk}.

\paragraph{Acknowledgements}

I would like to thank Nathan Benjamin, Scott Collier, Don Marolf and Eric Perlmutter for helpful discussions and comments, as well as the participants of the `Chaos and Order' program at the KITP, where this work was initiated, for stimulating discussions. I am grateful to be funded by a Len DeBenedictis Postdoctoral Fellowship, and for additional support received from the University of California. This research was supported in part by the National Science Foundation under Grant No. NSF PHY-1748958.

\section{The modular $\kernel$-transform}\label{sec:Stransform}

A key tool in our analysis is a formulation of modular invariance of $c>1$ theories directly as a property of the spectrum, with no direct reference to the partition function itself. This involves casting the $\tau \to -1/\tau$ S-transform of the torus into the form of an operation on the density of states, which we review and expand upon in this section.

\subsection{The modular S-matrix}

The traditional parameters in CFT -- central charge $c$ and dimensions $h,\bar{h}$ -- are not the most natural for describing Virasoro representation theory, so we will find it convenient to introduce different variables. In place of the central charge, we use $Q$ or $b$, defined by
\begin{equation}\label{eq:cParams}
	c=1+6Q^2,\quad Q=b+b^{-1},
\end{equation}
where we choose $0<b<1$ when $c>25$, so in particular $c\to\infty$ corresponds to $b\to 0$. For operator dimensions, we use $\alpha$ or $P$,
\begin{equation}\label{eq:hParams}
	h=\alpha(Q-\alpha) = \left(\tfrac{Q}{2}\right)^2+P^2,\quad \alpha = \frac{Q}{2} + i P,
\end{equation}
along with similar definitions for $\bar{P},\bar{\alpha}$.  Note that this splits the spectrum into two ranges, $h\geq \frac{c-1}{24}$ corresponding to real $P$, and $ h < \frac{c-1}{24}$, corresponding to imaginary $P$, in which case we will usually use $\alpha\in[0,\tfrac{Q}{2})$. These were referred to respectively as `continuum' and `discrete' ranges of operator dimensions in \cite{1811.05710}.
 We will sometimes refer to $P$ as a `momentum', a terminology from the linear dilaton and related theories, where it is a momentum in target space.

Now, the operator content of the theory is encoded by a density of primary states $\rho(P,\bar{P})$, which is a sum of delta functions supported at the locations of primary operators (in a theory with a discrete spectrum). In terms of this density, the partition function is
\begin{equation}\label{pfDensity}
	\pf(\tau,\bar{\tau}) = \int_0^\infty dP d\bar{P}\;  \chi_P(\tau)\bar{\chi}_{\bar{P}}(\bar{\tau}) \rho(P,\bar{P})= \int_{-\infty}^\infty \frac{dP}{2}\frac{d\bar{P}}{2}\; \chi_P(\tau)\bar{\chi}_{\bar{P}}(\bar{\tau})\rho(P,\bar{P}) ,
\end{equation}
where the latter equation defines $\rho$ as being even in $P$ and $\bar{P}$. The characters
\begin{equation}\label{chars}
	\chi_P(\tau) = \frac{e^{2\pi i \tau P^2}}{\eta(\tau)},\quad \bar{\chi}_{\bar{P}}(\bar{\tau}) = \frac{e^{-2\pi i \bar{\tau} \bar{P}^2}}{\eta(-\bar{\tau})},
\end{equation}
encode the contribution of all descendants of a nondegenerate Virasoro primary. In particular, this notation does not take into account the special case of the degenerate vacuum; to compensate, the density of states includes negative delta functions to subtract the null descendants. The vacuum contribution to $\rho(P,\bar{P})$ is $\rho_\id(P)\rho_\id(\bar{P})$, where\footnote{The delta functions supported at imaginary momentum may look unfamiliar; we clarify their definition in section \ref{ssec:maths} and appendix \ref{app:maths}.}
\begin{equation}\label{rhoVac}
	\rho_\id(P) = \left[\delta\left(P-i\tfrac{b^{-1}+b}{2}\right)-\delta\left(P-i\tfrac{b^{-1}-b}{2}\right) + (P\leftrightarrow -P)\right].
\end{equation}

Our results will be based on modular invariance of the partition function $\pf$, specifically invariance under the S-transform:
\begin{equation}
	\pf(\tau,\bar{\tau})=\pf(-1/\tau,-1/\bar{\tau})
\end{equation}
We will study this by asking the following question: given a particular set of primary states of given energy and angular momentum, if we construct a corresponding partition function, make a modular transformation, and decompose the result into Virasoro primaries in the `dual channel', what is the resulting spectrum? This procedure is enacted directly as a linear operator acting on the density $\rho$, the `modular S-matrix' $\kernel$, an extension of the object familiar from rational CFTs, for which $\kernel$ is a finite-dimensional matrix. The S-matrix we will derive was written down in \cite{hep-th/0101152} (in the context of open/closed duality for boundary CFT on the annulus), and applied in \cite{1407.6008} to construct candidate modular invariant spectra.

To find the S-matrix, begin with an alternative definition of the kernel $\kernel$, acting as a modular transform on the characters:
\begin{equation}
	\chi_{P}(\tau) = \int_{-\infty}^\infty dP'\; \chi_{P'}(-1/\tau)\;\kernel_{P'P} 
\end{equation}
We can decompose a partition function $\pf$ into characters in these two different channels, with spectrum $\rho$ in the `original' channel, and $\hat{\rho}$ in the `dual' channel. Suppressing dependence on barred variables for now, we have
\begin{align}
	\pf(\tau) &= \int_{-\infty}^\infty \frac{dP'}{2} \;\chi_{P'}(-1/\tau)\hat{\rho}(P')  \\
	&= \int_{-\infty}^\infty \frac{dP}{2}\; \chi_{P}(\tau)\rho(P)  \\
	&= \int_{-\infty}^\infty \frac{dP}{2}\; \int_{-\infty}^\infty dP'\; \chi_{P'}(-1/\tau)\;\kernel_{P'P} \;\rho(P) \\
	&= \int_{-\infty}^\infty \frac{dP'}{2} \;\chi_{P'}(-1/\tau)\left[\int_{-\infty}^\infty dP\; \kernel_{P'P}\;\rho(P) \right],
\end{align}
where we have used the S-matrix to transform the characters, before exchanging the order of integration.
We can now equate the first and final lines and drop the $P'$ integral\footnote{The map from the spectrum to the partition function is invertible, since the Laplace transform is injective.} to find that the modular transform acts as an integral transform with kernel $\kernel$, as desired (now restoring the barred sector):
\begin{equation}\label{eq:modularStrans}
	\hat{\rho}(P',\bar{P}') = \int_{-\infty}^\infty dP d\bar{P} \; \kernel_{P'P}\bar{\kernel}_{\bar{P}'\bar{P}} \;\rho(P,\bar{P})
\end{equation}

We now need only compute the kernel. In terms of momentum $P$, the characters \eqref{chars} are just Gaussians, so we are looking for a transform that maps Gaussians to Gaussians with inverse width. This is nothing but the Fourier transform, written with a slightly unfamiliar normalisation\footnote{In Mathematica, this is implemented by \texttt{FourierTransform}, using $\mathtt{FourierParameters}\to \{0,-4\pi\}$.}, having the following kernel:
\begin{equation}\label{eq:modS}
	\kernel_{PP'} = \sqrt{2}e^{-4\pi i PP'}
\end{equation}
Since we are always taking $\rho$ to be even and real, we get the same answer using $\sqrt{2}e^{4\pi i PP'}$, which is the kernel for the inverse transform. This means that, on the space of functions we are interested in, the Fourier transform squares to unity as expected for the modular S-transform\footnote{One natural perspective is to interpret $\kernel^2$, which maps $(P,\bar{P})\mapsto (-P,-\bar{P})$, as CPT conjugation. For example, for a free boson at radius $R$, we can take $P=n R^{-1}+w\frac{R}{2}$ and $\bar{P}=n R^{-1}-w\frac{R}{2}$ for momentum and winding modes $n,w\in\ZZ$, which are the charges of (anti-)holomorphic $U(1)$ currents, and flip sign under CPT.}. We could equivalently use a cosine transform, with kernel $\sqrt{2}\cos(4\pi PP')$.

An important special case is the transform acting on the vacuum density \eqref{rhoVac}:
\begin{align}\label{eq:rhoVacS}
	\rho_\id(P) &= \delta\left(P-i\tfrac{b^{-1}+b}{2}\right)-\delta\left(P-i\tfrac{b^{-1}-b}{2}\right) +\delta\left(P+i\tfrac{b^{-1}+b}{2}\right)-\delta\left(P+i\tfrac{b^{-1}-b}{2}\right)\nonumber \\
	&\implies \hat{\rho}_\id(P) = 4\sqrt{2}\sinh(2\pi b P)\sinh(2\pi b^{-1}P).
\end{align}

\subsection{Mathematics of the S-transform}\label{ssec:maths}

Some features of the density of states $\rho$, in particular delta functions supported at imaginary momentum from states with $h<\frac{c-1}{24}$, do not appear in conventional discussions of distributions and the Fourier transform. We must allow $\rho$ to live in a more general space of distributions than those commonly considered.

Distributions are defined as duals to some space of well-behaved test functions, that is as continuous linear functionals: a distribution $\rho$ maps a test function $\psi$ to a number $\langle\rho,\psi\rangle$, giving $\int dP \rho(P)\psi(P)$ when $\rho$ is an ordinary function. The Fourier transform is usually defined on the space of \emph{tempered distributions}, dual to  smooth and rapidly decaying (Schwartz) test functions (see \cite{strichartz2003guide}, for example). However, this does not encompass delta functions at imaginary points (a Schwartz function cannot be evaluated at complex values), nor exponentially growing densities like \eqref{eq:rhoVacS} (a Schwartz function $\psi$ may not decay rapidly enough for $\int \psi(P)e^{\lambda P}dP$ to converge). To enlarge our space of distributions sufficiently, we must choose a more restrictive space of test functions, which are entire analytic and decay faster than any exponential. For us, this space need only be large enough to contain the characters, as a function of $P$ for any fixed $\tau$ (that is, Gaussians).

For most of our purposes such mathematical details will not worry us, and it will be sufficient to manipulate the distributions na\"ively, but with the confidence that there is a rigorous theory underlying our results. Where more care is required, we will refer to appendix \ref{app:maths}, which contains further mathematical discussion.

\subsection{Cardy formulas}\label{sec:Cardy}

The modular S-matrix \eqref{eq:modS} determines how a single operator of dimension $h'$ contributes to the density of states at dimension $h$ after performing a modular S-transform. Fixing $h'$ and taking $h$ to be large, the S-matrix oscillates if $h'\geq \frac{c-1}{24}$ (real $P'$), but grows exponentially as $e^{2\pi P(Q-2\alpha')}$ for input operator dimension $h'<\frac{c-1}{24}$ (imaginary $P'$, real $\alpha'<\frac{Q}{2}$). The spectrum at large $P$ is therefore determined predominantly, after S-transform, by the operators of smallest $h'$.

For a unitary compact theory, this means that the density of states at large energy is dominated by the identity in the S-transform decomposition: for all states besides the vacuum, $\alpha'+\bar{\alpha}'$ is bounded below by a positive number, so their contribution is exponentially suppressed at large $P,\bar{P}$. The spectrum at large energy approaches the S-transform of the vacuum state\footnote{This argument is too fast, because it does not bound the contribution of a sum of infinitely many operators. Indeed, \eqref{eq:Cardy1} interpreted literally must be false, since $\rho$ is a sum of delta functions at integer spins. However, contributions from heavy states are oscillatory, so tend to cancel from an appropriately smeared density of states. Refining the argument to make more precise statements is an interesting problem which we do not attempt to address in this paper. See \cite{1904.06359} for related discussion.}:
\begin{equation}\label{eq:Cardy1}
\begin{gathered}
	\rho(P,\bar{P}) \sim  \hat{\rho}_\id(P) \hat{\rho}_\id(\bar{P}) \qquad 
		P,\bar{P}\to\infty \qquad (\text{Unitary, compact}) \\
		\rho_\id(P) \sim \sqrt{2} \exp\left[2\pi \sqrt{\tfrac{c-1}{6}\left(h-\tfrac{c-1}{24}\right)}\right]
\end{gathered}	
\end{equation}
 This is the Cardy formula for the asymptotic density of primary states \cite{Cardy:1986ie}, along with some correction terms\footnote{To express the density in terms of $h,\bar{h}$ or $\Delta,\ell$, a Jacobian for the change of variables must be included. This gives a logarithmic contribution to the entropy of primary states. To compare to the entropy of all states, such as in \cite{gr-qc/0005017,1205.0971}, we should add the `primary entropy' at $h_p\sim (1-c^{-1})h$ to the `descendant entropy' at $h_d \sim c^{-1}h$; the latter is given by the Hardy-Ramanujan asymptotic formula for partitions, and the allocation of energies in $h=h_p+h_d$ between primaries and descendants maximises the sum of these entropies. Formulas which are insensitive to spin, such as in \cite{1702.00423,1904.06359}, are obtained by integrating over $\ell$ at fixed $\Delta$.}. Further corrections to \eqref{eq:Cardy1} are exponentially suppressed in at least one of $P\sim\sqrt{h}$ and $\bar{P}\sim\sqrt{\bar{h}}$.

From this argument, it is a small extension to consider the large spin limit, where we take only $P\to\infty$, fixing $\bar{P}$. In this case, for the vacuum alone to dominate, it must be the only Virasoro primary with $h=0$, meaning that there is no extended current algebra. More precisely, we demand a twist gap, meaning that there is a positive lower bound on $h$ for all Virasoro primaries besides the vacuum. With this restriction on the theory in question, we obtain a large spin version of the Cardy formula:
\begin{equation}\label{eq:Cardy2}
	\rho(P,\bar{P}) \sim  \hat{\rho}_\id(P) \hat{\rho}_\id(\bar{P}) \quad 
		 P\to\infty,\, \bar{P} \text{ fixed} \qquad  (\text{Twist gap})
\end{equation}

In essence, this argument is equivalent to the `modular lightcone bootstrap' in \cite{1608.06241,1707.07717} (see also appendix B of \cite{1810.01335}), which considers the partition function with $\tau,\bar{\tau}$ independent and imaginary, in the limit $\tau\to i\infty$ with $\bar\tau$ fixed. However, we use the density of states directly, without reference to the partition function. This perspective is more useful when the energy and spin of interest never dominate the canonical ensemble, as can occur in a `near-extremal' limit where we take $\bar{P}\to 0$ sufficiently fast as $P\to\infty$: the large spin density \eqref{eq:Cardy2} at $P\bar{P}<\frac{c-1}{24}$ has greater free energy than the vacuum alone, so is always subdominant in the partition function to the identity (see \cite{1405.5137} in a context of large $c$ gapped theories). This makes it subtle to isolate the `continuum' piece of the spectrum in this regime from the identity by using the asymptotic behaviour of the partition function.

\subsection{The modular transform of large spin growth\label{sec:PW}}

It is familiar that the behaviour of a function at large parameters is related to the smoothness of its Fourier transform, by results going under the general name of Paley-Wiener theorems. For example, if $\rho$ decays as a power $P^{-2-n}$, then $\hat{\rho}$ is $n$ times continuously differentiable, and if $\rho$ decays as an exponential, then $\hat{\rho}$ has an analytic extension to a strip. Some further results in this tradition, involving distributions which grow at infinity, will be a useful heuristic for us to understand how different regimes of operator dimensions are related by modular transform.

Firstly, we have already met an exponentially growing distribution, the modular transform of the vacuum \eqref{eq:rhoVacS}, which behaves as $e^{2\pi Q |P|}$ in the $|P|\to\infty$ limit. Under Fourier transform, this is related to the fact that the vacuum has support at imaginary momentum $P$ (dual to the result that exponential decay transforms to analyticity in a strip):
\begin{equation}\label{eq:PW1}
	\rho(P) \text{ supported at} \Im P=\tfrac{Q}{2}-\alpha \iff \hat{\rho}(P)\approx e^{4\pi \left(\tfrac{Q}{2}-\alpha\right) P}
\end{equation}
In the large spin limit of \eqref{eq:Cardy2} ($P\to\infty$), this means that exponential growth comes from operators with low twist ($h<\frac{c-1}{24}$) in the modular transformed channel, and the fastest growth comes from the lowest twist. The asymptotic expansion at large spin is therefore organised by including operators in increasing order of twist.

Secondly, we will see that operators of low twist do not appear in isolation; rather, infinite families of multi-twist operators accumulate at a given twist, with density of states growing polynomially in spin. After Fourier transform, this shows up in the twist dependence of corrections to the large spin Cardy formula \eqref{eq:Cardy2}; the transform of a distribution of power law growth is supported on real $\bar{P}$, but with a particular singular part at $\bar{P}=0$. Roughly, since a power can be removed by repeated differentiation, the singularity at $\bar{P}=0$ is sufficiently weak that it is removed by multiplication by a power of $\bar{P}$.

A concrete example is the following S-transform, which will be used later:
\begin{equation}\label{eq:powerTransform}
	\rho_k(\bar{P}) = |\bar{P}|^{2k-1}\implies \hat{\rho}_k(\bar{P}) = (-1)^{k}2\sqrt{2}\frac{(2k-1)!}{(4\pi)^{2k}} \pv \frac{1}{\bar{P}^{2k}}
\end{equation}
We derive this result in appendix \ref{app:maths}, and give a precise definition of the distribution $\hat{\rho}_k$ (in particular the resolution of the singularity implied by the principal value symbol $\pv$). This leading singularity remains correct even for a density supported at discrete values of $\bar{P}$ (in particular, integer spins), as long as the integrated density (that is, the total number of operators up to some spin) is asymptotic to $\int_0^{\bar{P}} \rho_k = \frac{1}{2k}\bar{P}^{2k}$.

\section{The spectrum of multi-twist operators\label{sec:multispec}}

In this section, we discuss the multi-twist composite operators built from a light primary. We begin by reviewing the properties of double-twist operators in section \ref{sec:DT}, before using this to infer the universal properties of multi-twists in section \ref{ssec:multiTwist}. The key results for the sequel are \eqref{eq:ptwist} and \eqref{eq:largelDegen}, which give the asymptotic twist and number of operators (respectively) in multi-twist Regge trajectories. A reader impatient apply the modular bootstrap could take these results on trust and proceed directly to section \ref{sec:Smultitwist}.

 In section \ref{ssec:MTgravity}, we discuss a useful intuition and motivation for the multi-twist operators as multi-particle states in an AdS$_3$ dual. In \ref{sec:MTcorrections}, we discuss the non-universal properties of the multi-twist spectrum, subject to corrections which depend on details of the theory in question.

\subsection{Double-twist Regge trajectories from the fusion kernel\label{sec:DT}}

The existence of double-twist composite operators follows from crossing symmetry of a four point correlation function, much as the large spin Cardy formula follows from modular invariance. In this section we sketch the argument of \cite{1811.05710}, leading to the result \eqref{eq:DTspectrum} about the spectrum.

For any two primary operators $\op_1$, $\op_2$, consider a four-point correlation function $\langle \op_1\op_2\op_2\op_1\rangle$, and use the OPE to write it in terms of the basic data of the theory, namely the central charge, spectrum, and OPE coefficients. The same correlation function can be decomposed in several different ways, in particular the `S-channel', taking the OPE of $\op_1$ and $\op_2$, and the `T-channel', taking the OPE between pairs of identical operators. These two decompositions must arrive at the same result, which gives the crossing equations relating the S- and T-channel expansions.

In the previous section, modular invariance was recast as invariance of the density of states under the modular S-matrix integral transform \eqref{eq:modularStrans}. The crossing equation can be recast in an analogous way, giving an expression for S-channel spectral density (a sum of delta functions supported at the momenta $(P,\bar{P})$ of operators appearing in the OPE, weighted by OPE coefficients) as an integral transform acting on T-channel spectral density. The kernel of this transform is the \emph{fusion kernel}, or the Virasoro $6j$-symbol \cite{hep-th/9911110,arXiv:1202.4698}. Table \ref{table} summarises the analogous objects appearing in the partition function and the four-point function.
\begin{table}[h]
\centering
\begin{tabular}{c|c}
Partition function $\pf$ & Four-point function	$\langle \op_1\op_2\op_2\op_1\rangle$ \\
Density of states $\rho$ & Spectral density \\
Characters $\chi(\tau)$ & Conformal blocks \\
Modular invariance & Crossing equation \\
Modular S-matrix $\kernel$ & Fusion kernel \\
$\hat{\rho}_\id(P)$ (see \eqref{eq:rhoVacS}) & Identity fusion kernel
\end{tabular}
\caption{The analysis of partition functions and modular invariance is closely parallel to four-point functions and crossing\label{table}}
\end{table}

We highlight one qualitative difference: while for modular invariance the same data (the spectrum) appears in both channels, for crossing the S- and T-channel data are different (unless $\op_1=\op_2$), since an operator labelled by $p$ is weighted by OPE coefficients $C_{12p}^2$ or $C_{11p}C_{22p}$ in the S- and T-channel respectively (and one of these may vanish; most importantly, the identity appears in the T-channel but not S-channel).

For the modular bootstrap \eqref{eq:Cardy2}, we argued that at large spin, the contribution from the identity in the cross-channel is parametrically larger than the contribution from any operator of positive twist. The same argument applies for the four-point function: for a theory with a twist gap, the S-channel spectral density at large spin is dominated by the contribution from the identity in the T-channel, so is given by the identity fusion kernel to leading order in an asymptotic $\ell\to\infty$ expansion.

At this point the fusion kernel has a crucial new feature absent from the modular S-matrix: it can have support on some additional, discrete operator dimensions. The modular S-transform of the identity (or any single operator) gives a continuous density of states supported only on real $P$, that is dimensions $h\geq \frac{c-1}{24}$. The fusion transform of the identity includes a similar continuous spectral density for $h\geq \frac{c-1}{24}$, but also discrete delta function\footnote{These come from poles in the fusion kernel, so evaluating the residue gives a delta function, c.f.~analytic representations of distributions as described in appendix \ref{app:maths}.} contributions at a finite set of dimensions $h_m<\frac{c-1}{24}$, determined by the external operator dimensions. These are simple to express in the parameterisation \eqref{eq:hParams}, where they correspond to real $\alpha=\alpha_m<\frac{Q}{2}$:
\begin{equation}
	\alpha_m = \alpha_1+\alpha_2 + m b, \quad m\in \NN,\quad \alpha_m<\frac{Q}{2}
\end{equation}
Note that these discrete contributions only exist if $\alpha_1+\alpha_2<\frac{Q}{2}$, and for $m\geq 1$ only if $c>25$, in which case we have chosen $0<b<1$.

The upshot is that the twist dependence of the large spin S-channel spectral density -- the analogue of $\hat{\rho}_\id(P)$ in the large spin (here, $\bar{P}\to\infty$) Cardy formula \eqref{eq:Cardy2} -- includes delta functions supported at imaginary $P = \pm i(\tfrac{Q}{2}-\alpha_m)$. For each $\alpha_m$, there must be a family of operators (a Regge trajectory) labelled by spin $\ell$, approaching the corresponding twist as $\ell\to\infty$. This argument alone does not imply that there is an operator on this trajectory for every spin, but the Lorentzian inversion formula  \cite{1703.00278,1711.03816,1805.00098}, implying analyticity of spectral data in spin, supplies this missing link. The result is that, for each $m$ such that $\alpha_m<\frac{Q}{2}$, there are `double-twist' primary operators $[\op_1\op_2]_{m,\ell}$:
\begin{equation}\label{eq:DTspectrum}
	\text{For all }\ell,\quad \exists [\op_1\op_2]_{m,\ell}, \text{ with } \alpha_{m,\ell}\to \alpha_1+\alpha_2+mb \text{ as }\ell\to\infty.
\end{equation}
All $\ell$ here means every integer spin starting at the sum of the spins of the component operators\footnote{We are here organising by representations of the connected part of the conformal group, so spin can be any integer; including parity (if a symmetry of the theory) would combine positive and negative spins in a single representation. There are also double-twist Regge trajectories for states spinning the opposite way, with $\ell\leq \ell_1+\ell_2$ and $\bar{\alpha}$ approaching the universal values as $\ell\to-\infty$ (this double counts the operators with $\ell=\ell_1+\ell_2$ at the start of the trajectories). The precise counting here is based on expectations from MFT.}, $\ell\geq \ell_1+\ell_2$ (with the possible exception of small spins, for which the inversion formula may not converge); if $\op_1=\op_2$ we have only even spins.

In the $c\to\infty$ limit with dimensions of external operators held fixed, the number of double-twist trajectories (the range of allowed $m$) is of order $c$, and the asymptotic twists of each trajectory are $h_{m,\ell}\sim h_1+h_2+m + O(c^{-1})$. The double-twist spectrum approaches that of mean field theory (generically requiring $\ell\gg c$ to reach the large spin regime).

The rate at which the Regge trajectories approach their asymptotic twist $\alpha_m$ at large spin (the `anomalous twist' $\gamma_{m,\ell}$) is determined by the operator of lowest twist $\bar{\alpha}_t$ exchanged in the T-channel besides the identity:
\begin{equation}\label{eq:DTanom}
	\gamma_{m,\ell} = \alpha_{m,\ell}-(\alpha_1+\alpha_2+mb)\sim \gamma_m e^{-2\pi \bar{\alpha}_t \sqrt{\ell}}
\end{equation}
The coefficient $\gamma_m$ (given in \cite{1811.05710}) is proportional to OPE coefficients $C_{11t}C_{22t}$.

\subsection{Extending to multi-twists\label{ssec:multiTwist}}

Given the existence of double-twist operators, it is only a small extension to deduce that there are also higher composite `multi-twist' operators -- though counting them is slightly more challenging. In this section, we describe the most naive proposal for the spectrum, and motivate its correctness (at least to leading order in large spin) and universality.

To establish the existence of multi-twist operators and determine certain properties, it is enough to recursively apply the argument for double-twist operators. Here we will start with a single operator $\op$; the extension to multiple species is straightforward. For the first step, constructing triple-twists, simply run the argument of section \ref{sec:DT} taking one of the external operators to be a double-twist, $\op_1=[\op\op]_{m_\text{DT},\ell_\text{DT}}$, with $\op_2=\op$. The resulting operators $\left[[\op\op]_{m_\text{DT},\ell_\text{DT}}\op\right]_{\ell, m}$ are the triple-twists. The large spin bootstrap is valid if the spin of the double-twist we started with is large ($\ell_\text{DT}\gg 1$) and the spin of the resulting triple-twist is much larger still ($\ell-\ell_\text{DT}\gg 1$); then the twist of the composite must approach the universal value $3\alpha + (m_\text{DT}+m)b$.\footnote{This requires a mild assumption about OPE coefficients of double-twists at large $\ell_\text{DT}$.} We can then repeat the exercise taking $\op_1$ to be a triple-twist and so forth, to give $p$-fold composites approaching the following twists at large spin:
\begin{equation}\label{eq:ptwist}
	\alpha\sim \alpha_{p,m} = p\alpha + m b \qquad p,m\in \NN, \quad \alpha_{p,m}<\tfrac{Q}{2}.
\end{equation}
This argument applies until $p$ is large enough that $p\alpha >\frac{Q}{2}$, and we can no longer separate distinct, discrete Regge trajectories from the $h>\frac{c-1}{24}$ large spin `continuum'. Note that not all multi-twist operators will approach these universal twists at large spin; however, in \ref{sec:MTcorrections} we argue that most of them do, in a sense to be made precise momentarily.

This recursive argument alone is not, however, sufficient to correctly count the multi-twists. Enumerating all operators constructed in this way will overcount, because the same multi-twist may be built in several ways, in particular due to the Bose symmetry permuting copies of $\op$.
At this point, we make the very natural suggestion that the multi-twist spectrum can be regarded (at least to leading order at large spin) as a deformation of mean field theory (MFT, also called a generalised free field). MFT is not a true CFT, but a set of correlation functions on the plane solving the crossing equations. The correlators of $\op$ are Gaussian, given by Wick contractions using a conformal two-point function, and the stress tensor decouples\footnote{Usually MFT is described as having no stress tensor. An alternative point of view is that the stress tensor exists in its own decoupled free sector, but normalised by $\langle T(0)T(1)\rangle=1$ rather than by Ward identities, so it is absent from the OPE of other operators at $c=\infty$. This is convenient for counting states, because we can continuously take the formal $c\to\infty$ decoupling limit.} ($c\to\infty$). The spectrum of `multi-trace' operators in MFT is given by a non-interacting Bose gas built from $\op$ and its global descendants. We can write the $p$-trace operators as normal-ordered products,
 \begin{align}\label{eq:MFTMT}
 	&:\!  (\partial\bar{\partial})^{m_1}\bar{\partial}^{\ell_1}\op\cdots (\partial\bar{\partial})^{m_p}\bar{\partial}^{\ell_p}\op\!:, \\
 	\text{with twist }& h_{p,m}=p h_\op+m, \; m=\sum_{i=1}^p m_i\quad \text{and spin }\ell = p \ell_\op + \sum_{i=1}^p \ell_i. \nonumber
 \end{align}
 These operators are mostly $\sl$ descendants, and only particular linear combinations are primary. These $\sl$ primaries are in fact full Virasoro primaries, with Virasoro descendants built by dressing \eqref{eq:MFTMT} with $T$, $\bar{T}$.\footnote{Regarding MFT as a $c\to\infty$ limit where $T$ decouples, the states are those of a free Bose gas with three families of particle species, $T$ and $\bar{T}$ as well as $\op$, and their global descendants. The pairs $L_{-n},L_{n}$ for $n\geq 2$ in the Virasoro algebra (appropriately rescaled) become creation and annihilation operators for $\partial^{n-2}T$.}

 We enumerate the multi-trace states of MFT in appendix \ref{app:counting}. The main result we will require is the large spin degeneracy of primaries at any given twist (that is, fixed $m$). For each $p,m$, the total number of primaries with spin at most $\ell$ grows as follows:
\begin{equation}\label{eq:largelDegen}
	N_{p,m}(\ell)\sim  \frac{(p+m-2)!}{m!(p-2)!} \frac{1}{(p-1)! p!}\; \ell^{p-1}
\end{equation}
For example, for $p=2$ we get $N_{2,m}(\ell)\sim \frac{\ell}{2}$ for every $m$, as expected for a state on each double-twist Regge trajectory for each even spin. Our assumption is that this formula remains true in our generic, finite $c$ CFT for multi-twist Virasoro primary operators with $\alpha\sim \alpha_{p,m}$.

It is straightforward to extend these results to include several independent `single-twist' operators, and construct mixed multi-twists from them; we can also allow for fermionic operators. These generalisations are discussed in appendix \ref{app:counting}, and the relevant results introduced when required in section \ref{sec:Smultitwist}.

\subsection{Multi-twists as multi-particle states in AdS\label{ssec:MTgravity}}

A strong motivation and intuition for the multi-twist spectrum comes from an interpretation in an AdS dual, which is useful even at finite $c$ and when no local, semiclassical dual exists \cite{1212.3616,1403.6829,1212.4103}.

In this context, it is simplest to use the state-operator correspondence to talk not of local primary operators, but the spectrum of primary states on $S^{d-1}$ (for a CFT in $d$ dimensions). Each such state corresponds to an excitation in global AdS$_{d+1}$, with `centre of mass' wavefunction in the ground state. For Virasoro primaries, the boundary gravitons are also in their lowest energy state, which in particular means that the boundary stress-tensor expectation value is constant. Taking global descendants boosts the configuration so the excitation orbits AdS, confined by the potential induced by the cosmological constant. Taking large spin, the excitation orbits a large proper distance from the centre of AdS.

Multi-twist states now come from making several such excitations, each with large angular momentum, such that they are well separated. A `cluster decomposition' principle in AdS then suggests that the excitations become independent in the limit of large separation, forming a non-interacting Fock space with appropriate statistics. Note that this does not require the excitations to be `elementary' or localised in any sense; it may occupy any region of finite size. At small spin, there is no longer a parametric separation between the excitations, so we may not neglect interactions, and so the spectrum depends on details of the theory and excitations.

It is not obvious that this should make sense for a generic CFT, for which there is no obvious bulk description, and if it exists it will not be local on AdS scales (for example, Planck or string scales will be of order $\ell_\text{AdS}$). Nonetheless, the bootstrap analysis shows that a notion of locality emerges at very large distance, on scales parametrically larger than the curvature of AdS. The notion of a gravitational dual is therefore useful in far more generality than might have been expected, in the context of a large spin analysis corresponding to a bulk long distance expansion. For certain quantities, any CFT is describable by a low-energy bulk effective field theory with an AdS scale cutoff (this is related to, but distinct from the `effective conformal theory' notion of \cite{1007.2412}).

The above description is only really valid in $d>2$, in which case all unitary interactions fall off at long distance. The situation for $d=2$ is complicated by infinite-range interactions mediated by massless particles, which cannot be neglected even at very large separation. This includes the effect of gravity, by which a localised excitation induces a conical defect, which remains important at all distances. Such interactions correspond to currents in the CFT, with vanishing twist $\bar{h}=0$; our assumption of a twist gap ensures that it suffices to include only the effect of gravity. While we do not currently have a satisfactory bulk description of the gravitational effects at finite $c$, the Virasoro bootstrap \cite{1811.05710} shows how to account for them in the spectrum.

\subsection{Corrections to the multi-twist spectrum\label{sec:MTcorrections}}

The previous section identified the leading order large spin spectrum of multi-twists: there are families of operators approaching each of the twists in \eqref{eq:ptwist} as $\ell\to\infty$, and their number is given by \eqref{eq:largelDegen} up to order $\ell^{-1}$ corrections. This is the only data we can determine universally, that is without additional detailed information about other operators and their coupling to $\op$.

Double-twist Regge trajectories have anomalous twists \eqref{eq:DTanom}, which are small at large spin; the recursive construction extends this to estimate anomalous twists of many of the multi-twist operators. As long as each of the $p-1$ spins added at successive stages of the construction is large (an order one fraction of the final total spin $\ell$, say), we can apply the large spin bootstrap at each stage, and the multi-trace anomalous dimension is suppressed exponentially in $\sqrt{\ell}$.

There are certainly multi-twist operators to which the above does not apply. The anomalous twists of double-twist operators need not be suppressed in any sense at finite $\ell$; furthermore, when we build higher multi-traces, large anomalous twists will appear even in the large spin spectrum, accumulating at values of the twist different from those in \eqref{eq:ptwist} as $\ell\to\infty$. However, such operators are relatively few. Roughly speaking, the spins added at each stage in the recursive construction correspond to the $\ell_i$ in \eqref{eq:MFTMT} (except that one $\ell_i$ is dropped, since we are forming primaries), and the $\ell^{p-1}$ growth in \eqref{eq:largelDegen} comes from the number of ways to choose $(p-1)$ component spins summing to less than $\ell$ (up to a factor from permutation symmetry). For most of these, each $\ell_i$ contributes an order one fraction to the total $\ell$, and hence the corresponding operator has small anomalous twist. The proportion of decompositions for which at least one of the $\ell_i$ is less than any fixed number, allowing for large anomalous twist, is of order $\ell^{-1}$. Hence, the number of $p$-twist operators with large anomalous dimension grows only as $\ell^{p-2}$, smaller than the total \eqref{eq:largelDegen} by a factor of $\ell^{-1}$.

Finally, we expect that there can be fewer multi-twists than anticipated from the Bose gas of MFT, because the inversion formula \cite{1703.00278} does not converge for all spins. For scalar $\op$, the double-twist trajectories need only extend to spin two, excluding scalar double-twists (though we should comment that this applies to $\sl$ primaries, and it is not entirely clear how it extends to Virasoro primaries). For higher multi-twists, this can lead to a number of `missing operators', which is again $\ell^{-1}$ suppressed relative to \eqref{eq:largelDegen}.

\section{The S-transform of multi-twist operators}\label{sec:Smultitwist}

In this section, we combine the large spin spectrum of multi-twist operators, \eqref{eq:ptwist} and \eqref{eq:largelDegen}, with modular invariance in the framework of section \ref{sec:Stransform}. From this, we will recover corrections to \eqref{eq:Cardy2} for the `near extremal' regime of large $h$ with $\bar{h}$ close to $\frac{c-1}{24}$, which we interpret as shifts of the `extremality bound'.

Before the calculations, we motivate why the S-transform of multi-twists should tell us about the near extremal spectrum, using the Paley-Wiener type results of section \ref{sec:PW}. Starting with the operator $\op$ of lowest twist, we are looking at some of the pieces of the spectrum with largest $|\Im P|$; after modular transform, these become some of the most important terms in the $P\to\infty$ large spin expansion, from \eqref{eq:PW1}. We have also seen that, for a given twist, the simplest and most universal piece of the multi-twist spectrum is the leading power law growth of states with spin; this leading power of $\bar{P}$ leads, after S-transform, to the most singular function of twist from \eqref{eq:powerTransform}, and hence the terms that are most important for small $\bar{P}$.
\begin{equation}
	\boxed{\begin{aligned}
		\text{Lowest twist } \alpha \\
		\text{Largest power } \ell\to\infty
	\end{aligned}} \xrightarrow{\text{S-transform}} \boxed{\begin{aligned}
		\text{Largest for } P\to\infty \\
		\text{Most singular } \bar{P}\to0 
	\end{aligned}}
\end{equation}
  While the higher multi-twist (large $p$) operators have greater twist, and hence result in contributions that are more suppressed in the large spin $P\to\infty$ limit, they also grow faster in number and so these terms are enhanced for small $\bar{P}$. Their importance is thus highlighted in a particular combined limit $P\to\infty$, $\bar{P}\to 0$, which focusses on the `near extremal' spectrum.

\subsection{Shift in extremality bound from universal multi-twists}

Begin by taking the $p$-twist operators with asymptotic twist \eqref{eq:ptwist}, and write the growth of states \eqref{eq:largelDegen} at large $\ell$ in terms of a density\footnote{The actual density is a sum of delta functions at integer spins, but the smooth density suffices to capture the singularity of the Fourier transform; see appendix \ref{app:maths}.} in $\bar{P}$:
\begin{equation}\label{eq:Reggedensities}
\begin{aligned}
	\rho_{p,m}&(P,\bar{P}) = \frac{(p+m-2)!}{m!(p-2)!} \frac{2}{(p-2)! p!} |\bar{P}|^{2p-3} \\
	\times & \left[\delta\left(P+i\left(\tfrac{Q}{2}-p\alpha-m b\right)\right)+\delta\left(P-i\left(\tfrac{Q}{2}-p\alpha-m b\right)\right)\right]
\end{aligned}
\end{equation}
Using the results of section \ref{sec:Stransform}, in particular \eqref{eq:Cardy1}, it is simple to find the spectrum corresponding to the modular transform of these states (keeping only the term with exponential growth large $P$ and dropping the decaying term, and leaving the $\pv$ symbol implicit):
\begin{align}
	\hat{\rho}_{p,m}(P,\bar{P}) &= \frac{2(p+m-2)!}{(p-2)!^2p!m!}\frac{(-1)^{p-1}2\sqrt{2}(2p-3)!}{(4\pi)^{2(p-1)}} \frac{1}{\bar{P}^{2(p-1)}}\times\sqrt{2}e^{4\pi P\left(\tfrac{Q}{2}-p\alpha-m b\right)}\nonumber \\
	&= 8 e^{2\pi Q P} (2\pi \bar{P})^2 \binom{\tfrac{1}{2}}{p} \left(2\pi \bar{P} e^{2\pi\alpha P}\right)^{-2p} (-1)^m \binom{1-p}{m} e^{-4\pi m b P}   \label{eq:rhohatpm}
\end{align}
A nice feature of this formula is that it has a useful interpretation even for $p=0$ and $p=1$, which was not evident from the starting point \eqref{eq:Reggedensities}. At $p=1$, only the $m=0$ term is nonzero, and it gives $4e^{2\pi (Q-2\alpha) P}$, which is the small $\bar{P}$ limit of the S-transform of the operator $\op$ alone (keeping only the term with growing exponential in $P$). The $p=0$ terms are nonzero for $m=0,1$, and give the leading term in a small $\bar{P}$ expansion of the S-transform of the vacuum \eqref{eq:rhoVacS} (again keeping only growing exponentials in $P$); the $m=1$ term serves to subtract the left-moving null descendants of the vacuum.

Now for each of the finitely many $p,m$ such that $p\alpha+m b<\frac{Q}{2}$, the density of states includes a contribution $\hat{\rho}_{p,m}$. Further, it is likely that all other contributions are either more suppressed at large spin $P\to\infty$ (coming from operators of higher twist), or less singular at $\bar{P}\to 0$ (coming from slower growth in spin for a given twist). In this case, the expressions \eqref{eq:rhohatpm} for $\hat{\rho}_{p,m}$ are the most important terms in a near-extremal asymptotic expansion, taking $P\to\infty$ and $\bar{P}\to 0$ (fixing $\bar{P}e^{2\pi \alpha P}$, perhaps).\footnote{It is likely that some of the $m\geq 1$ trajectories will have a larger twist for given asymptotic growth than other operators excluded from the analysis, and hence give terms which are less important than some omitted corrections in any limit.}

 However, individual terms in the expansion do not have a particularly satisfactory interpretation, particularly since they give densities which are singular as $\bar{P}\to 0$. To remedy this, we will sum the terms \eqref{eq:rhohatpm} into a simple closed form. This requires extrapolating \eqref{eq:rhohatpm} to all integers $p,m\geq 0$, and extending to an infinite sum. The terms we add are all exponentially small at large spin, so are not even meaningful as corrections to a discrete spectrum, which becomes a smooth density only in an approximation where there are parametrically many relevant states. A conservative interpretation of the resummed density is simply a convenient repackaging of the near-extremal expansion, valid to the highest order which still contributes exponentially growing terms.

Once we have extended \eqref{eq:rhohatpm} to include all $m,p\in\NN$ (including $p=0,1$ for the vacuum and $\op$ itself), the sums over $m$ and then $p$ are simply binomial series, and we can write the answer as follows:
\begin{equation}\label{eq:sumspectrum}
	\sum_{p,m=0}^\infty \hat{\rho}_{p,m}(P,\bar{P}) \stackrel{?}{=} (8\pi)^2 e^{2\pi b^{-1} P}\sinh(2\pi b P)  |\bar{P}|  \sqrt{\bar{P}^2+\frac{1}{(2\pi)^{2}}\frac{e^{-4\pi\alpha P}}{1-e^{-4\pi b P}}}
\end{equation}
However, this manipulation -- as indicated by the interrogative equality -- is somewhat too na\"ive, since we were taking an infinite sum over singular distributions and treating them as ordinary functions. A more careful analysis, described in appendix \ref{app:maths}, shows that the correct sum is not simply the smooth function of $\bar{P}$ in \eqref{eq:sumspectrum}, but a distribution with additional support at imaginary values of $\bar{P}$, extending to the point where the square root vanishes. Rewriting \eqref{eq:sumspectrum} as a density in terms of twist $\bar{h}=\frac{c-1}{24}+\bar{P}^2$ by including the Jacobian factor $\frac{d\bar{P}}{d\bar{h}}$, we find a density
\begin{equation}
	\rho(P,\bar{P})\frac{d\bar{P}}{d\bar{h}} \sim 2\sqrt{2}(2\pi)^2\hat{\rho}_\id(P) \sqrt{\bar{h}-\frac{c-1}{24}+\frac{1}{(2\pi)^{2}}\frac{e^{-4\pi\alpha P}}{1-e^{-4\pi b P}}},
\end{equation}
which is supported where the argument of the square root is positive. The corrections from this universal piece of the multi-twist spectrum therefore have a simple interpretation, providing a spin-dependent shift to the `extremality bound' $\bar{h}_\text{extr}$, the edge of the effective continuum of states appearing at large spin:
\begin{equation}\label{eq:extrShift1}
	\boxed{\bar{h}_\text{extr} - \frac{c-1}{24} \sim -\frac{1}{(2\pi)^2}\frac{e^{-4\pi\alpha P}} {1-e^{-4\pi b P}}}
\end{equation}
This captures only the leading terms in a large spin expansion of the extremality bound; in fact, some of the terms that come from expanding the denominator (from $m\geq 1$ trajectories) may be less important than the omitted corrections.

We should be clear that the argument here does not `prove' \eqref{eq:extrShift1}; we have simply derived a finite number of terms in an asymptotic expansion. Indeed, in a generic CFT there is no sharp notion of $\bar{h}_\text{extr}$ at finite spin, so it is unclear how to precisely define \eqref{eq:extrShift1} beyond consistency with an asymptotic expansion. The interpretation of the series as a shift of the $\sqrt{}$ edge of the spectrum is nonetheless extremely compelling, in particular since it requires a very rigid structure of the expansion: a single operator contribution (the $p=1$ term here) uniquely determines the leading singularity at all subsequent orders (all $m$ for all $p>1$). The truncation of the expansion after a finite number of terms is an expected feature when $\bar{h}_\text{extr}$ is only an approximate notion at large spin. We will see further evidence for for this interpretation in section \ref{sec:SCMI}, in a semiclassical $c\to\infty$ limit where there is a sharp extremality bound for spins of order $c$. In that context, the shift of $\bar{h}_\text{extr}$ appears more directly and explicitly.

We have so far included only a particular leading order set of terms. We expect further corrections to have two effects. Firstly, we can have additional terms in the asymptotic expansion of $\bar{h}_\text{extr}$, either from additional operators or including more details about multi-twists. Secondly, we can have terms which are less singular in the $\bar{P}\to 0$ limit, which will not affect the extremality bound, but instead correct the functional form $\rho_\id(\bar{P})$ of the density of states in a large spin expansion.

\subsection{Multiple operators}\label{sec:multiOps}

The result of our analysis came from considering the modular transform of a single light operator and the large spin multi-twist descendants constructed from it. Here, we show that adding more light operators, along with the multi-twists constructed from all possible combinations thereof, is consistent with the same interpretation and changes the result simply by summing the corrections to $\bar{h}_\text{extr}$.

Start with operators $\op_i$, with dimensions given by $(\alpha_i,\bar{\alpha}_i)$, where $i$ runs from $1$ to $N$. The multi-twist trajectories are labelled by a particle number $p_i$ for each species, as well as $m$. The asymptotic degeneracies, computed in appendix \ref{app:counting}, are given by \eqref{eq:largelDegen} for composites of $p=\sum p_i$ operators of a single type, times a multinomial coefficient:
\begin{equation}
\begin{gathered}
\alpha\sim \sum_i p_i \alpha_i + m b,\\
	N_{\{p_i\},m}(\ell)\sim
	\frac{p!}{\prod_{i=1}^N p_i!} \frac{(p+m-2)!}{m!(p-2)!} \frac{1}{(p-1)! p!}\; \ell^{p-1}
\end{gathered}
\end{equation}
For example, the double-twist Regge trajectories built out of distinguishable particles ($p_1=p_2=1$) number a factor of two larger than for identical particles, since there are double-twist primaries at both even and odd spin.

Now the derivation continues as before, with an extra step of summing over all $p_i$ with $\sum_i p_i=p$ (a multinomial series), before summing over $p$. The result is that the contributions from different operators inside the square root in the density of states \eqref{eq:sumspectrum} simply add:
\begin{equation}
	\bar{h}_\text{extr} \sim \frac{c-1}{24} -\frac{1}{(2\pi)^2} \sum_{i=1}^N\frac{ e^{-4\pi\alpha_i P}} {1-e^{-4\pi b P}}
\end{equation}

\subsection{Fermions}

It is straightforward to repeat the analysis for a fermionic operator $\op$. The main difference is the statistics of multi-traces, forming a Fermi gas (more details are in appendix \ref{app:counting}). However, this does not change the count at leading order in large spin, so \eqref{eq:largelDegen} remains valid.

In addition to the statistics, we also have a choice of boundary conditions when fermions are involved. In a path integral, this is a choice of antiperiodic (Neveu-Schwarz) or periodic (Ramond) identification going round the two independent cycles of the torus. For the Hilbert space interpretation, the spatial boundary condition determines which states (living in NS or R Hilbert spaces) are being counted, and the Euclidean time boundary condition determines whether fermions get counted with signs (inserting $(-1)^F$ in the trace for periodic).

Choosing the multi-trace operators to always belong to the NS sector Hilbert space, as does the vacuum, we can perform our modular bootstrap for either choice of the Euclidean time boundary conditions (NS-NS or NS-R); after S-transform, this means we are counting NS or R states without $(-1)^F$ insertion (NS-NS or R-NS). For the NS sector, since the large spin growth of degeneracies is unaltered, the result \eqref{eq:extrShift1} for Bosons carries over immediately to Fermions:
\begin{equation}
	\delta\bar{h}_\text{extr}  = -\frac{1}{(2\pi)^2}\frac{e^{-4\pi\alpha P}} {1-e^{-4\pi b P}} \quad (\text{NS sector, F or B})
\end{equation}

 For the R sector, the insertion $(-1)^F$ in the trace becomes an inclusion of $(-1)^p$ before summing over \eqref{eq:rhohatpm}. This simply swaps the sign inside the square root \eqref{eq:sumspectrum}, meaning that fermion multi-trace contributions shift the threshold in the opposite way for the R sector:
 \begin{equation}
	\delta\bar{h}_\text{extr}  = \pm \frac{1}{(2\pi)^2}\frac{e^{-4\pi\alpha P}} {1-e^{-4\pi b P}} \quad (\text{R sector, } \pm \leftrightarrow \substack{F\\ B})
\end{equation}

\subsection{Anomalous dimensions\label{sec:anomalous}}

The analysis of this section so far used multi-twist operators with exactly linear Regge trajectories, that is with twist (or $\alpha$) independent of spin (or $\bar{P}$). This is true only in a leading order approximation at large spin, and the operators receive anomalous dimensions as discussed in section \ref{sec:MTcorrections}.
 Here, we characterise how this correction contributes to the S-transformed spectrum for the double-twist trajectories, and argue that it is of subleading importance for the extremality bound; something similar should be true for higher multi-twists, though we have not explicitly checked.

For double-twist families, including the anomalous dimension \eqref{eq:DTanom} modifies the asymptotic density \eqref{eq:Reggedensities} as follows (where a conjugate delta function has been dropped):\footnote{The operators live at even spins $\ell=\bar{P}^2-P^2$. To convert this into a density in $\bar{P}$ needs a Jacobian factor $\frac{d\ell}{d\bar{P}}$, which gets modified by anomalous twists. The additional correction from this factor is suppressed by an extra $\log P$.}
\begin{equation}
	\rho_{2,m}(P,\bar{P})\sim |\bar{P}| \delta\Big(P+i\big(\tfrac{Q}{2}-2\alpha-m b-\gamma_m e^{-2\pi \bar{\alpha}_t |\bar{P}|}-\cdots \big)\Big)
\end{equation}
We now approximate how this deviates from the spectrum \eqref{eq:rhohatpm} without anomalous twists, after taking the S-transform. The Fourier transform in $P$ is straightforward from the delta function, so we are left with the following integral for the $\bar{P}$ transform:
\begin{equation}
	\delta\hat{\rho}_{2,m}(P,\bar{P})\sim  2e^{4\pi\left(\tfrac{Q}{2}-2\alpha-m b\right)P} \Re\int^\infty d\bar{P}' \, e^{ 4\pi i \bar{P} \bar{P}'} \bar{P}' \left[\exp\left({-4\pi \gamma_m e^{-2\pi\bar\alpha_t\bar{P}'} P}\right)-1\right]
\end{equation}
We leave the lower bound of the integral ambiguous, since we are only looking at contribution from $\bar{P}'\to\infty$ where the large spin expansion of anomalous twist is valid. Low spin operators are discussed at the end of the section.

Now, for any fixed $P$, the integrand decays exponentially as $\bar{P}'\to\infty$, so the resulting transform is an analytic function of $\bar{P}$. This means in particular that we do not have a singularity, like the $\bar{P}^{-2}$ in \eqref{eq:rhohatpm}. However, it is less clear what happens in a simultaneous limit taking $P\to\infty$ along with $\bar{P}\to 0$; for this, it is easiest to make a change of variable which factors out the $P$ dependence:
\begin{equation}
	x = 2\pi\bar{\alpha}_t \bar{P}'-\log P\implies e^{-2\pi\bar\alpha_t \bar{P}'} P = e^{-x}
\end{equation}
This is designed to remove the $P$ dependence from the square bracket in the integrand. The other factors then give a $\log P$ enhancement, along with an oscillatory exponential in $\frac{2}{\bar{\alpha}_t} \bar{P}\log P$. What remains is the Fourier transform of an exponentially decaying function of $x$, and hence an analytic function of $\bar{P}$:
\begin{equation}
	\delta\hat{\rho}_{2,m}(P,\bar{P})\sim  e^{4\pi\left(\tfrac{Q}{2}-2\alpha-m b\right)P} \log P \Re \left[e^{\frac{2i}{\bar{\alpha}_t}  \bar{P}\log P} \times (\text{analytic in }\bar{P})\right]
\end{equation}
This correction competes with $\hat{\rho}_{2,m}$ if $\log P \gtrsim \bar{P}^{-2}$, and hence is unimportant in the regime of the shift of the extremality bound \eqref{eq:extrShift1}, with $\bar{P}$ scaling as an exponential of $P$.

Note that if we perform the change of variable to $x$ and integrate over $x>0$, this corresponds to $\bar{P}'\gtrsim \log P$,  which includes double-twists with spin of order $(\log P)^2$ and greater. We can include lower spins in the above argument by a simple bound on their contribution to the S-transform density of states, coming from the number of such operators times the largest modular S-matrix for any one of them. The largest possible contribution comes from the operator with largest negative anomalous dimension, and hence lowest twist $\alpha_\text{min}$. Multiplying the number of neglected operators (of order $(\log P)^2$), we have a bound $(\log P)^2 e^{4\pi (\frac{Q}{2}-\alpha_{\min})P}$. This is of course nonsingular at $\bar{P}=0$, and is a small correction as long as $\alpha_\text{min}>\alpha$ (so the single operator we started with indeed had lowest twist).

To go to higher orders in the large $P$ expansion of $\bar{h}_\text{extr}$ \eqref{eq:extrShift1}, we can first treat any double-traces of particularly low twist (large negative anomalous dimension) as independent operators, using the result of section \ref{sec:multiOps}. Even if anomalous twists were entirely absent, there would still be corrections to the number of multi-twist operators, beyond the leading order in spin power \eqref{eq:largelDegen}. A reasonable guess is that these corrections take the form of terms added to \eqref{eq:extrShift1}, starting at order $e^{-8\pi\alpha P}$ from including a constant term in the number of double-traces. We will see an example of additional terms of this form in the semiclassical limit.

\section{Semiclassical AdS$_3$ gravity}\label{sec:bulk}

We now explore the gravitational interpretation of the results of section \ref{sec:Smultitwist}, as a quantum shift of the extremality bound of rotating BTZ black holes when AdS$_3$ gravity is coupled to matter.

\subsection{BTZ and the extremality bound}

We begin by discussing the extremality bound for rotating black holes in pure gravity, and its relation to the large spin bound $\bar{h}_\text{extr}>\frac{c-1}{24}$.

A general stationary axisymmetric metric in three dimensions can be written in the form
\begin{equation}\label{eq:aximetric}
	ds^2 = -n(r)^2 f(r) dt^2+\frac{dr^2}{f(r)}+ r^2 (d\phi+k(r)dt)^2.
\end{equation}
Solving the vacuum Einstein equations with cosmological constant $\Lambda = -1$ (choosing units with $\ell_\text{AdS}=1$) gives the BTZ metric \cite{hep-th/9204099,gr-qc/9302012},
\begin{equation}
	f(r) = \frac{(r^2-r_+^2)(r^2-r_-^2)}{r^2}, \quad k(r) = \frac{r_+r_-}{r^2},\quad n(r)=1,
\end{equation}
and the parameters $r_\pm$ are determined in terms of the mass and angular momentum (classically) by
\begin{equation}
	M = \frac{r_+^2+r_-^2}{8 G_N}, \quad J = \frac{r_+ r_-}{4 G_N}.
\end{equation}
The mass here is defined such that empty AdS$_3$ has energy $-\frac{1}{8G_N}$ (again, classically). In terms of scaling dimensions of the corresponding CFT operators, we have $M = h+\bar{h}-\frac{c-1}{12}$ and $J=\ell = h-\bar{h}$, so
\begin{equation}
	h = \frac{c-1}{24}(1+(r_+ + r_-)^2), \quad \bar{h} = \frac{c-1}{24}(1+(r_+ - r_-)^2),
\end{equation}
where we have also used the Brown-Hennaux central charge $c=\frac{3}{2G_N}$ \cite{Brown:1986nw}, up to a shift of $c$ to $c-1$, explained momentarily. There are particularly simple expressions in terms of the momentum variables \eqref{eq:hParams}:
\begin{equation}
	P = \frac{Q}{2}(r_+ + r_-), \quad \bar{P} = \frac{Q}{2}(r_+ - r_-).
\end{equation}

The causal structure is determined largely by the zeros of $f$. The fastest outgoing null geodesics follow $\frac{dr}{dt} = n(r)f(r)$, so there is a horizon at the largest value of $r$ for which this vanishes ($r=r_+$ for BTZ). Under the cosmic censorship assumption that the singularity at $r=0$ is shrouded by a horizon, $f$ must have a positive real root, which implies the extremality bound
\begin{equation}
	\text{No naked singularity} \iff h,\bar{h}\geq\frac{c-1}{24}.
\end{equation}
This is saturated by the extremal black hole $r_-=r_+$, so $P=Q r_+$ and $\bar{P}=0$. In fact, BTZ above the extremality bound and empty AdS$_3$ exhaust all\footnote{This classification is up to diffeomorphism, but some `large' diffeomorphisms are physical, acting as asymptotic symmetries; this is interpreted as dressing with a coherent superposition of Virasoro descendants.}
 exterior\footnote{Other solutions exist, but they are always isometric to BTZ in any region outside a horizon, causally connected to an asymptotic boundary.} solutions of pure AdS$_3$ gravity without naked singularities.

The shift of $c$ to $c-1$ is a one-loop effect from metric fluctuations. It can be interpreted simply as a $-\frac{1}{12}$ contribution to $M$ from the Casimir energy of gravitons.\footnote{There is a similar Casimir energy for empty AdS, which should be interpreted as a one-loop renormalisation of the Brown-Hennaux relation \cite{Brown:1986nw} between $c$ and $G_N$. AdS$_3$ has classical energy $-\frac{c_\text{bare}}{12}$ with $c_\text{bare}= \frac{3}{2G_N}$, and graviton Casimir energy $-\frac{1}{12}-1$, where the subtraction of unity is due to the invariance of the vacuum under $L_{-1},\bar{L}_{-1}$; this adds up to $-\frac{c}{12}$ with $c=c_\text{bare}+13$ (see \cite{1808.03263} for a useful perspective).} A less direct way to see this is from a Euclidean partition function for fluctuations around BTZ, which is a modular transform of Euclidean AdS$_3$ with thermal identifications. The one-loop graviton partition function on the latter (which is exact to all orders in perturbation theory) gives the CFT vacuum character \cite{0712.0155,0804.1773}, with ground state energy determined as $-\frac{c}{12}$ by conformal invariance. Performing the modular transform to go back to BTZ, we find a spectrum supported on $h,\bar{h}\geq\frac{c-1}{24}$ (with density of primary states \eqref{eq:rhoVacS}). While this closely resembles the CFT derivation of the Cardy formula, it is a calculation purely in semiclassical gravity and requires no CFT dual.

\subsection{Including a scalar field at one loop}

To include the effect of matter, take the simplest example of a free scalar minimally coupled to Einstein gravity:
\begin{equation}
	S = \frac{1}{16\pi G_N}\int d^3 x \sqrt{-g} (R+2) - \int d^3 x \sqrt{-g} \left[\tfrac{1}{2}(\nabla\Phi)^2+\tfrac{1}{2} m^2\Phi^2\right]
\end{equation}
To one-loop order, any weakly interacting theory of a scalar coupled to the metric can be brought to this form; for example, a curvature coupling $R\Phi^2$ can be absorbed by a Weyl transformation of the metric. Integrating out the scalar at one loop sources the Einstein equation at order $G_N$ with the expectation value of the stress tensor, giving a quantum correction to the geometry. We will find the range of mass and angular momentum for which the backreacted black hole has a horizon.

Our analysis generalises previous work for a massless conformally coupled scalar with `transparent' boundary conditions\footnote{The conformal coupling gives an effective mass in AdS corresponding to $m^2=-\frac{3}{4}$ after an appropriate Weyl rescaling. The transparent boundary conditions are not conformally invariant: the resulting propagator is a linear combination of those for $\Delta=\tfrac{1}{2}$ and $\Delta=\tfrac{3}{2}$.}, for which the stress tensor expectation value was computed in \cite{gr-qc/9308032}. In this special case, the linearised Einstein equations were solved in \cite{1608.05366,1902.01583}.

Computation of the expectation value $\langle T_{ab}\rangle$ is relatively straightforward since BTZ is locally isometric to AdS$_3$, obtained as a quotient $\text{AdS}_3/\Gamma$ by a subgroup $\Gamma$ of its $SO(2,2)$ isometry group. For BTZ, $\Gamma \simeq \ZZ$ as a group, generated by a single element, which acts to identify the angular coordinate as $\phi\sim \phi+2\pi$. The Hartle-Hawking state of the free scalar $\Phi$ on such a geometry is characterised as the Gaussian state on which the one-point function $\langle\Phi\rangle$ vanishes, and the two-point function is given by the method of images, using the AdS$_3$ propagator $G_\Delta$ (discussed in a moment):
\begin{equation}
	\langle \Phi(x)\Phi(x') \rangle_{\text{AdS}_3/\Gamma} = \sum_{\gamma\in \Gamma} G_\Delta(x,\gamma\cdot x')
\end{equation}

Variation of the action gives the classical stress-tensor:
\begin{equation}
	T_{ab} = \nabla_a\Phi \nabla_b \Phi-\tfrac{1}{2}g_{ab}((\nabla\Phi)^2+m^2\Phi^2)
\end{equation}
The expectation value $\langle T_{ab}\rangle$ is therefore given by a differential operator acting on the two-point function, with an appropriate regularisation to take the limit of coincident points. The description as a quotient makes regularisation straightforward; we may simply drop the term in the sum over images when $\gamma$ is the identity. This is equivalent to regularising the divergence and adding a counterterm which renormalises the cosmological constant, chosen such that the renormalised stress tensor expectation value vanishes in pure AdS$_3$.
\begin{equation}\label{eq:Timagesum}
	\langle T_{ab}\rangle(x) =  \sum_{\gamma\in \Gamma-\id}\left. \left[\nabla_a \nabla'_b -\tfrac{1}{2}g_{ab}(g^{cd}\nabla_c \nabla'_d+m^2)\right] G_\Delta(x,x')\right|_{x'=\gamma\cdot x}
\end{equation}

The AdS$_3$ propagator $G_\Delta$ solves the equation of motion with a $\delta$-function source at coincident points,
\begin{equation}
	(\Box-\Delta(\Delta-2)) G_\Delta(x,x') = \delta^{(3)}(x,x'), \qquad m^2=\Delta(\Delta-2),
\end{equation}
and obeys appropriate boundary conditions at infinity; in the coordinates \eqref{eq:aximetric}, the solution decays as $r^{-\Delta}$, with no component scaling as $r^{\Delta-2}$. The propagator is a function only of proper distance $s$ between $x$ and $x'$, obeying
\begin{equation}
	G_\Delta''(s) +2\coth(s) G_\Delta'(s) = \Delta(\Delta-2) G_\Delta(s)
\end{equation}
for $s>0$, with explicit solution
\begin{equation} \label{eq:propagator}
	G_\Delta(s) = \frac{1}{2\pi} \frac{e^{-\Delta s}}{1-e^{-2s}}.
\end{equation}

We now take the expression \eqref{eq:Timagesum} and write $G$ as a function of proper distance $s_\gamma(x,x') = s(x,\gamma\cdot x')$, eliminating the mass term using the equation of motion satisfied by $G_\Delta$:
\begin{multline}\label{eq:TEV}
	\langle T_{ab}\rangle(x) =  \sum_{\gamma\in \Gamma-\id} \Big[\left(\nabla_a s_\gamma \nabla'_b s_\gamma -\tfrac{1}{2}g_{ab}(g^{cd}\nabla_c s_\gamma \nabla'_d s_\gamma+1)\right) G_\Delta''(s_\gamma) \\
	 + \left(\nabla_a \nabla'_b s_\gamma -\tfrac{1}{2}g_{ab}(g^{cd}\nabla_c \nabla'_d s_\gamma + 2\coth s_\gamma) \right) G_\Delta'(s_\gamma) \Big]
\end{multline}
Written in this form, the stress tensor is conserved as an identity for any function $G_\Delta$, and we will not need to use any information about the propagator until the very end.

The final required ingredient is an expression for the proper distance $s_\gamma(x,x')$. A convenient way to calculate this is to express the geometry as a quotient of the $SL(2,\RR)$ group manifold. For extremal BTZ, an explicit form is
\begin{equation}
	g(t,r,\phi) = \frac{1}{\sqrt{2}} \begin{pmatrix}
		1 & r_+(t-\phi) \\
		0 & 1
	\end{pmatrix} \cdot \begin{pmatrix}
		\frac{r_+}{\sqrt{r^2-r_+^2}} & \frac{r_+}{\sqrt{r^2-r_+^2}} \\
		-\frac{\sqrt{r^2-r_+^2}}{r_+} & \frac{\sqrt{r^2-r_+^2}}{r_+}
	\end{pmatrix} \cdot \begin{pmatrix}
		e^{r_+(t+\phi)} & 0 \\
		0 & e^{-r_+(t+\phi)}
	\end{pmatrix},
\end{equation}
with metric $ds^2 = -\det(dg)$, and the quotient acts by imposing $2\pi$ periodicity on $\phi$. We can now use a simple expression for the proper distance in $SL(2,\RR)$ (see \cite{1412.0687}, for example),
\begin{align}\label{eq:sexBTZ}
	\cosh s &= \tfrac{1}{2} \Tr(g^{-1}(t_1,r_1,\phi_1) g(t_2,r_2,\phi_2)) \nonumber \\ 
	&= \frac{1}{2}\left(\sqrt{\frac{r_1^2-r_+^2}{r_2^2-r_+^2}}+ \sqrt{\frac{r_2^2-r_+^2}{r_1^2-r_+^2}} \right) \cosh(r_+(\Delta \phi+\Delta t)) \\ &\qquad + \frac{\sqrt{(r_1^2-r_+^2)(r_2^2-r_+^2)}}{2r_+} (\Delta\phi-\Delta t)\sinh(r_+(\Delta \phi + \Delta t)),  \nonumber
\end{align}
where $\Delta\phi = \phi_2-\phi_1$, $\Delta t = t_2-t_1$. The distances $s_\gamma(x,x')$ for different preimages in the quotient are obtained by adding integer multiples of $2\pi$ to $\Delta\phi$.

With all these ingredients, we could simply push ahead and solve the linearised Einstein equations with source \eqref{eq:TEV}. While this is possible (in fact, the solution is algebraic in $G_\Delta$ and its derivatives for a general function $G_\Delta$), we can extract the information of interest much more simply using conservation laws.

\subsection{Conserved quantities}

 We are interested in the relationship between the variation of the metric at infinity -- namely the shift of energy and angular momentum due to the matter source -- and at the horizon, where we impose `cosmic censorship' in the form of existence of the horizon. In pure gravity (or any diffeomorphism invariant theory, on-shell), such a relationship is provided by the first law of black hole thermodynamics $dM-\Omega dJ=T dS$. We will make use of a formulation of the first law in Einstein gravity derived from the covariant phase space methods of Wald et.~al.~\cite{gr-qc/9307038,gr-qc/9911095,gr-qc/9403028} (see \cite{hep-th/0503045} for application to asymptotically AdS spacetimes), allowing for an arbitrary conserved source, which adds an additional term given by an integral of the stress tensor over a Cauchy surface.\footnote{I would like to thank Don Marolf for helpful discussions regarding this section.}

We review the relevant constructions of the covariant phase space formalism in appendix \ref{app:Wald}. The most important object for us is the `Hamiltonian variation' $\delta\mathbf{H}_\xi$ corresponding to the vector field $\xi$, which is a $(d-1)$-form (in $d+1$ dimensional spacetime) depending on the background metric and a variation. If $\xi$ generates an asymptotic symmetry, then the integral of $\delta\mathbf{H}_\xi$ on a spatial surface at infinity gives the variation of the corresponding ADM conserved quantity $H_\xi$.

 Im the case that both the background and the variation solve the equations of motion, and if $\xi$ is a Killing field for the background, then $\delta\mathbf{H}_\xi$ is a closed form. If we generalise to allow for any variation, for us sourced by the one-loop stress tensor of the scalar, then $d\delta\mathbf{H}_\xi$ is proportional to the linearised equations of motion. The result is the following conservation equation:
\begin{equation}\label{eq:conservation}
d\delta\mathbf{H}_\xi = - \star(T_{ab}\xi^a dx^b)
\end{equation}

We now integrate this equation over a Cauchy surface $\Sigma$, for us a slice of constant $t$ between the horizon of BTZ and the AdS boundary, and use Stokes' theorem ($n$ is the unit vector normal to $\Sigma$):
\begin{equation}
\delta H_\xi =  \int_{S^{1}_\infty}	 \delta\mathbf{H}_\xi = \int_{S^1_{\text{Hor.}}} \delta\mathbf{H}_\xi + \int_\Sigma d^2x \sqrt{\gamma}\, T_{ab}n^a\xi^b
\end{equation}
To evaluate the boundary terms at infinity and on the horizon bifurcation surface, we use explicit expressions for $\delta\mathbf{H}_{\xi}$ derived for Einstein gravity in appendix \ref{app:Wald}. For stationary axisymmetric variations around BTZ, in the gauge \eqref{eq:aximetric}, the expressions for the two Killing fields $\partial_t$ and $\partial_\phi$ are
\begin{align}
\delta \mathbf{H}_{\partial_t} &=  \frac{1}{16\pi G_N}\left(-\delta f(r) - \frac{2r_+^2r_-^2}{r^2}\delta n(r) - r_+r_-r\delta k'(r)\right)	d\phi + \cdots \\
\delta \mathbf{H}_{\partial_\phi} &=  \frac{1}{16\pi G_N}\left(-2r_+ r_-\delta n(r)- r^3\delta k'(r)\right)	d\phi + \cdots
\end{align}
where we have kept only the $d\phi$ component. Imposing asymptotically AdS boundary conditions, the integrals at infinity are variations of conserved quantities as expected:
\begin{align}
\delta H_{\partial_t} &= -\frac{1}{8G_N} \lim_{r\to\infty}\delta f(r) = \delta M \\
\delta H_{\partial_\phi} &= -\frac{1}{8G_N} \lim_{r\to\infty}r^3\delta k'(r) = \delta J
\end{align}

Now we choose the particular linear combination of $\partial_t$ and $\partial_\phi$ which is normal to the horizon, the field $\xi_K$ for which the event horizon is a Killing horizon:
\begin{equation}
\xi_K = \partial_t-\Omega \partial\phi, \quad \Omega = \frac{r_-}{r_+}	
\end{equation}
For this choice, the horizon integral is
\begin{equation}
	\int_{S^1_{\text{Hor.}}} \delta\mathbf{H}_{\xi_K} = -\frac{1}{8G_N}\delta f(r_+),
\end{equation}
which is precisely the data which determines whether a horizon is present for variations around the extremal geometry.

\subsection{The modified extremality bound}

The extremality bound gives the set of conserved quantities for which the corresponding geometry has an event horizon. For variations around extremal BTZ, this is determined by the sign of $\delta f(r_+)$: for $\delta f(r_+)>0$, $f+\delta f$ does not have a root near $r=r_+$, and so the singularity at $r=0$ becomes causally connected to the boundary. The linearised extremality bound is therefore $\delta f(r_+)\leq 0$, which we can rewrite  using the conservation equations:
\begin{equation}
 \int_{S^1_\text{Hor.}}\delta\mathbf{H}_{\xi_K} \geq 0 \implies \delta M-\delta J \geq 	\int_\Sigma d^2x \sqrt{\gamma}\, \langle T_{ab}\rangle n^a\xi^b
\end{equation}
It remains only to evaluate the integral, using \eqref{eq:TEV} for the stress tensor expectation value.

For extremal BTZ and sources respecting the symmetries, we have
\begin{align*}
	\int_\Sigma d^2x \sqrt{\gamma}\, \langle T_{ab} \rangle n^a\xi^b &= 2\pi \int_0^\infty dr \frac{r^3}{(r^2-r_+^2)^2}\left[\langle T_{tt}\rangle+\frac{r_+^2}{r^2}\langle T_{\phi\phi}\rangle-\frac{r^2+r_+^2}{r^2} \langle T_{t\phi}\rangle\right] \\
	&= \sum_{n=1}^\infty \int_{r_+}^\infty dr\; \left[A_n(r) G''_\Delta(s_n(r))+B_n(r) G'_\Delta(s_n(r))\right],
\end{align*}
where in the second line we have substituted using \eqref{eq:TEV}, and combined terms corresponding to an element of the quotient group and its inverse. From \eqref{eq:sexBTZ}, $s_n(r)$ is the proper length of a geodesic to and from a point at radius $r$, wrapping $n$ times round the horizon: 
\begin{equation}
	\cosh s_n(r) = \cosh (2 \pi  n r_+) + \frac{n \pi  \left(r^2-r_+^2\right)}{r_+}\sinh (2 \pi  n r_+)
\end{equation}
The expressions for $A_n$ and $B_n$ are rather complicated, but can be written in a form allowing an enormous simplification of the integral:
\begin{align}
A_n(r) &= \frac{(\cosh s_n(r)-\cosh (2 \pi  n r_+)) (2 \pi  n r_+ \cosh s_n(r)-\sinh (2 \pi  n r_+))}{\pi n^2  \sinh (2 \pi  n r_+) \sinh s_n(r)}	 s_n'(r) \nonumber \\
B_n(r) &= \frac{d}{dr}\left(\frac{A_n(r)}{s_n'(r)}\right) + \frac{1}{\pi n^2} s_n'(r)
\end{align}
Now we may integrate $A_n G_\Delta''$ by parts, which cancels the first term in the above expression for $B_n$. What remains is a total derivative, expressible in terms of $G_\Delta$ at the horizon (requiring only that $G_\Delta(s)$ goes to zero as $s\to\infty$):
\begin{equation}
\int_\Sigma d^2x \sqrt{\gamma}\, \langle T_{ab} \rangle n^a\xi^b = - \sum_{n=1}^\infty \frac{1}{\pi n^2} G_\Delta(2\pi n r_+)	
\end{equation}
Only at this stage do we need the explicit expression \eqref{eq:propagator} for the propagator, with which we write the one-loop bound in terms of the twist $\bar{h}$:
\begin{equation}\label{eq:extrResult}
\boxed{
	\bar{h} \geq \bar{h}_\text{extr} = \frac{c-1}{24} -\sum_{n=1}^\infty\frac{1}{(2\pi n)^2}\frac{e^{-2\pi n r_+\Delta}}{1-e^{-4\pi n r_+}}}
\end{equation}
Matching parameters to CFT variables in the semiclassical limit, we have $r_+ \sim b P$ and $\Delta\sim 2b^{-1}\alpha$. The $n=1$ term of the sum, which dominates in the large spin limit, matches \eqref{eq:extrShift1}. From linearity of the one-loop calculation, the contributions to $\delta\bar{h}_\text{extr}$ from multiple fields will add.

We expect this result to be valid for finite $r_+$ in the large $c$ limit, with weakly interacting bulk fields; in particular, for small black holes $r_+\sim c^{-1}$ the loop corrections are not suppressed.

\section{Semiclassical bootstrap}\label{sec:SCMI}

We now reproduce the result \eqref{eq:extrResult} of the gravity calculation from CFT considerations, using modular invariance in a large central charge limit, assuming that the spectrum of light states is given by MFT, or a Bose gas of free particles in AdS$_3$.

In general, the partition function for a gas of noninteracting Bosons is given by
\begin{equation}
	\pf_\text{Bose}(\beta) = \exp\left(\sum_{n=1}^\infty\frac{1}{n}\pf_\text{SP}(n \beta)\right),
\end{equation}
where $\pf_\text{SP}$ is the partition function for single-particle states. For a free particle in AdS$_3$, including independent left- and right-moving temperatures\footnote{This is the grand canonical ensemble for spin, with $\beta=\beta_L+\beta_R$ and real chemical potential $\mu= \frac{\beta_L-\beta_R}{\beta_L+\beta_R}$.}, $\pf_\text{SP}$ is a character of the global conformal $\sl\oplus \sl$ algebra:
\begin{equation}
	\pf_\text{Bose}(\beta_L,\beta_R) = \exp\left(\sum_{n=1}^\infty \frac{1}{n}\frac{q^{h_\Phi n}\bar{q}^{\bar{h}_\Phi n}}{(1-q^n)(1-\bar{q}^n)}\right),\: q=e^{-\beta_L}, \;\bar{q}=e^{-\beta_R}
\end{equation}
If we include the Virasoro descendants, which accounts for states with gravitons, as well as the shift from the ground state Casimir energy, we find the following contribution to the CFT partition function:
\begin{equation}\label{eq:pfBose}
	\pf(\beta_L,\beta_R)\supseteq \frac{e^{\frac{Q^2}{4}(\beta_L+\beta_R)}}{\eta(i\frac{\beta_L}{2\pi})\eta(i\frac{\beta_R}{2\pi})}(1-e^{-\beta_L})(1-e^{-\beta_R})\pf_\text{Bose}(\beta_L,\beta_R)
\end{equation}
The factors of $1-e^{-\beta}$ cancel an overcounting of descendants, since $\pf_\text{Bose}$ already includes global descendants (generated by $L_{-1},\bar{L}_{-1}$). The exception is the vacuum, but in that case the same factors are required to subtract the null descendants ($L_{-1},\bar{L}_{-1}$ annihilate the vacuum).

We now take a modular S-transform, and find the density of states corresponding to the gas of free particles in the dual decomposition. For real $\beta_{L,R}$, the density of primary states is related to the partition function by a two-variable Laplace transform:
\begin{gather}
	\int dE_L dE_R \,\rho(E_L,E_R)e^{-\beta_L E_L-\beta_R E_R} = \eta(i\tfrac{\beta_L}{2\pi})\eta(i\tfrac{\beta_R}{2\pi}) \pf\left(\tfrac{(2\pi)^2}{\beta_L},\tfrac{(2\pi)^2}{\beta_R}\right)\nonumber \\
	E_L = h-\tfrac{c-1}{24} = P^2,\quad E_R = \bar{h}-\tfrac{c-1}{24} = \bar{P}^2 \label{eq:pfLaplace}
\end{gather}
Here, $\rho$ denotes the density of Virasoro primary states with respect to the left- and right-moving energies $E_{L,R}$ defined above, so differs by a factor of $4P\bar{P}$ from the density used in earlier sections. Putting equations \eqref{eq:pfBose} and \eqref{eq:pfLaplace} together allows us to extract the density of states by inverse Laplace transform. In particular, the resulting ratios of $\eta$-functions simplify using the modular property $\eta(i\tfrac{\beta}{2\pi})= \sqrt{\tfrac{2\pi}{\beta}}\eta(i\tfrac{2\pi}{\beta})$.

If we ignore the gas of particles, setting $\pf_\text{Bose}$ to unity in \eqref{eq:pfBose}, the left- and right-moving pieces factorise, and we can perform the inverse Laplace transforms in closed form:
\begin{equation}
	\mathcal{L}^{-1}\left[ \sqrt{\tfrac{2\pi}{\beta}}e^{\frac{(\pi Q)^2}{\beta}}\left(1-e^{-\frac{(2\pi)^2}{\beta}}\right)\right](E) = \sqrt{\tfrac{8}{E}}\sinh\left(2\pi b\sqrt{E}
	\right)\sinh\left(2\pi b^{-1}\sqrt{E}\right)
\end{equation}
This amounts to an alternative derivation of the density S-dual to the vacuum state \eqref{eq:rhoVacS}.

Including the factor of $\pf_\text{Bose}$, it is not so simple to perform the inverse Laplace transforms, but our purposes do not require an exact result. We are interested in the near-extremal spectrum, which means taking $E_L$ to be of order $c$, but $E_R$ to be small. The inverse Laplace transform in the left-moving variables can then be performed by saddle-point in the large $c$ limit, and for the right-moving dependence we only need $\pf_\text{Bose}$ for large $\beta_R$:
\begin{equation}\label{eq:ZlargebetaR}
	\log\pf_\text{Bose}\left(\frac{(2\pi)^2}{\beta_L},\frac{(2\pi)^2}{\beta_R}\right) \sim \beta_R \sum_{n=1}^\infty \frac{1}{(2\pi n)^2} \frac{q^{h_\Phi n}}{1-q^n},\quad q=e^{-(2\pi)^2/\beta_L}
\end{equation}
The linear dependence on $\beta_R$ in the exponential simply gives a shift of $E_R$ in the resulting spectral density. The left-moving temperature is evaluated at the saddle-point $\beta_L \sim \frac{\pi}{b\sqrt{E_L}}$, which gives  $q = e^{-4\pi b \sqrt{E_L}}$. Writing in terms of momentum $P$ and twist $\bar{h}$, we find the following shift in the edge of the spectrum:
\begin{equation}
\boxed{
	\bar{h}_\text{extr} -\frac{c-1}{24} \sim - \sum_{n=1}^\infty \frac{1}{(2\pi n)^2} \frac{e^{-4\pi n h_\Phi b P}}{1-e^{-4\pi n b P}}}
\end{equation}
Taking $\Phi$ to be a scalar, $h_\Phi =\bar{h}_\Phi = \frac{\Delta_\Phi}{2}$, this precisely matches the result \eqref{eq:extrResult} of the extremality bound from the quantum corrected geometry.

If we have multiple species of particle, the shift in the extremality bound is simply a sum of the constituents: the Bose partition functions for each species simply multiply, so their contributions add in \eqref{eq:ZlargebetaR}.

The $n=1$ term matches \eqref{eq:extrShift1} in the appropriate semiclassical limit. The additional terms are due to the inclusion of the full multi-trace spectrum, rather than just the leading piece at large spin. If we were to include non-gravitational interactions to shift the energies of the multi-particle states, we expect the $n=1$ term to remain invariant, but higher terms in the sum to receive corrections.

\appendix

\section{The mathematical appendix\label{app:maths}}

In this section, we will describe the main mathematical ideas for defining and manipulating the distributions we encounter, and the Fourier transform. We will not attempt to be mathematically rigourous, in particular leaving out details of topologies, completeness, convergence and so forth.

\subsection{Distributions}

First, recall Schwartz's definition of distrubutions. The idea is to generalise the notion of function, by noting that an integrable function $\rho$ is characterised by its integrals against some well-behaved `test functions' $\psi$, $\langle \rho,\psi\rangle:=\int^\infty_{-\infty} dP \, \rho(P)\psi(P)$. This defines a linear functional on the space of test functions, uniquely determining $\rho$ almost everywhere (if that space is large enough). But this notion is now simple to generalise to nice linear functionals that do not correspond to any integrable function; for example, evaluation at a point $P_0$ defines the Dirac distribution $\delta_{P_0}$: $\langle\delta_{P_0},\psi\rangle := \psi(P_0)$.

A standard choice for the space of test functions is the Schwartz space, consisting of smooth functions such that all derivatives decay faster than any polynomial (leading to the `tempered distributions'). We will require a smaller space of test functions, which allows us to define a correspondingly larger space of distributions.\footnote{Another standard choice takes test functions to be smooth with compact support. We require analytic test functions, which can never have compact support, except for $0$.}

To formulate a well-behaved theory of distributions requires an extra technical ingredient, namely a topology on the space of test functions, and distributions are required to be continuous functionals with respect to this topology. We will not address this aspect here, but it will be important for a more complete and rigorous treatment.

We can define various operations on distributions by formal manipulations of the heuristic $\langle \rho,\psi\rangle \text{``=''} \int^\infty_{-\infty} dP \, \rho(P)\psi(P)$. Derivatives of distributions are defined by a formal integration by parts; multiplication by sufficiently nice functions $f$ and the Fourier transform $\rho\mapsto \hat{\rho}$ are defined by formal exchange of the order of integration:
\begin{align*}
	\langle \rho',\psi\rangle &:= -\langle \rho,\psi'\rangle\\
	\langle f\rho,\psi\rangle &:= \langle \rho,f\psi\rangle\\
	\langle \hat{\rho},\psi\rangle &:= \langle \rho,\hat{\psi}\rangle
\end{align*}
These implicitly require that the operation on the right maps test functons to test functions; the last of these in particular requires the space of test functions to be invariant under Fourier transform, which applies, for example, to the space of Schwartz functions.

As indicated in the text, to define the distributions we are interested in requires a more restricted space of test functions than is standard. Some properties we might like of the test functions (and their topology) are the following:
\begin{enumerate}
	\item Test functions are entire analytic.
	\item Test functions decay faster than any exponential on the real axis.
	\item The Fourier transform maps test functions to test functions.
	\item The Gaussians $\chi(\tau)$ are test functions, and their linear span is dense in the subspace of even functions.
\end{enumerate}
The Gaussians in the last example are, for us, the characters $\chi(\tau):P\mapsto\chi_P(\tau)\propto e^{2\pi i \tau P^2}$ for $\tau$ in the upper half-plane. Requiring them to be dense in the chosen topology ensures that a distribution $\rho$ can be determined uniquely from its partition function $\pf(\tau) = \langle\rho,\chi(\tau)\rangle$. Given the third requirement, the first two are somewhat redundant; the Fourier transform of a super-exponentially decaying function is entire.

The salient example of a distribution is a delta function supported at $P$, denoted $\delta_P$, and defined by $\langle \delta_P,\psi\rangle = \psi(P)$. While this is familiar from the more conventional spaces of distributions and tempered distributions, we can take $P$ to be any complex number, since we take the test functions to have entire analytic extensions. More generally, we can define distributions which integrate the test function over curves or regions in $\CC$. The other important property of the test functions is their decay rate, which enlarges the space of distributions to include exponential growth.

\subsection{A Fourier transform}

We here take the Fourier transform of the distribution $\rho_k(P)= 2k |P|^{2k-1}$, which we use in the text to describe the growth of degeneracies of multi-twist operators at large spin. We can also define this for $k=0$ by taking limit of these distributions, finding $\rho_0(P)=2\delta(P)$.

A simple way to work out the Fourier transform of such a distribution is by differentiating and using the usual property of the Fourier transform under derivatives, recursively on $k$:
\begin{align*}
	\rho_k(P) &= 2k|P|^{2k-1} \\
	\rho_k''(P) &= 2k(2k-1)\rho_{k-1}(P)\quad (k\geq 1) \\
	-(4\pi P)^2\hat{\rho}_k(P) &= 2k(2k-1)\hat{\rho}_{k-1}(P)\quad (k\geq 1) \\
	\rho_0(P) &= 2\delta(P)\\
	\hat{\rho}_0(P) &= 2\sqrt{2}
\end{align*}
If we now just go ahead and na\"ively divide by the factors of $P^2$, we get the following for the Fourier transform:
\begin{equation}
	\rho_k(P) = 2k|P|^{2k-1}\implies \hat{\rho}_k(P) = \frac{(-1)^{k}2\sqrt{2}(2k)!}{(4\pi)^{2k}} \pv \frac{1}{P^{2k}}
\end{equation}
This is subtle for two related reasons. Firstly, the distribution is singular, so needs a `regularistion' to properly define it (hence the inclusion of the principal value symbol $\pv$). Secondly, the devision by $P^2$ adds ambiguities supported at the origin (which must be $\delta$-functions and derivatives), which we must fix. The result is correct with the following definition of integration against an even test function $\psi$ (and vanishing for odd test functions):
\begin{equation}\label{pvDist}
	\left\langle \pv \frac{1}{P^{2n}},\psi(P)\right\rangle := \int \frac{dP}{P^{2n}}\left(\psi(P)-\sum_{k=0}^{n-1}\frac{\psi^{(2k)}(0)P^{2k}}{(2k)!}\right) \quad (\psi \text{ even})
\end{equation}
This means that we subtract enough terms of the Taylor expansion of the test function for the integrand to be finite at $P=0$, before integrating. To prove this is correct, and in particular show that this definition does not leave out any $\delta$-function pieces, it suffices to check with a Gaussian test function. Note that, because we're subtracting the `zero mode' of the test function $\psi(0)$, adding these terms shouldn't be interpreted as changing the total number of states; for example, adding $-\pv\frac{1}{P^2}$ removes some states at positive $P$, but adds them back in at $P=0$ in some sense.

\subsection{Analytic representations}

Since we are using analytic test functions, there is a nice alternative definition of the principal value distribution \eqref{pvDist} encountered above, by deforming the integral into the complex plane away from the singularity at $P=0$:
\begin{equation}
	\left \langle \pv \frac{1}{P^{2k}},\psi\right\rangle = \frac{1}{2}\int_{\RR+i \epsilon} dP \frac{\psi(P)}{P^{2k}} + \frac{1}{2}\int_{\RR-i \epsilon} dP \frac{\psi(P)}{P^{2k}}
\end{equation}

This is an example of a more general tool, an \emph{analytic representation} of a distribution (closely related to Sato's notion of hyperfunctions). Namely, for a given distribution $\rho$, there is a function $\Omega$ analytic everywhere except the real axis, such that $\rho$ is the discontinuity of $\Omega$ across the real axis, in the following sense:
\begin{equation}
\lim_{\epsilon\to 0} \int_{-\infty}^\infty dP \,[\Omega(P-i\epsilon)-\Omega(P+i\epsilon)] \psi(P) = \langle \rho,\psi\rangle
\end{equation}
In the instance above, we have $\Omega(P) = -\tfrac{1}{2} P^{-2k} \sgn{\Im P}$. Another nice example is $\Omega(P) = \frac{1}{2\pi i P}$, corresponding to $\delta(P)$. For more background, details, examples and applications see \cite{bremermann1965distributions}.

For us, having analytic test functions allows us to generalise the notion of analytic representations, in particular giving us representations of distributions with support away from the real axis. We only require that $\Omega$ is analytic outside a strip, for sufficiently large $|\Im P|$. We then say that $\Omega$ is an analytic representation of $\rho$ if 
\begin{equation}
 \int_\Gamma dP\, \Omega(P) \psi(P) = \langle \rho,\psi\rangle \qquad \forall \psi,
\end{equation}
where $\Gamma$ is a contour running from left to right in the region of analyticity in the lower half-plane, and similarly from right to left in the upper half-plane.

For a square integrable function, an analytic representation can be determined by a Cauchy integral, which is convolution with the analytic representation of the $\delta$-function:
\begin{equation}
	\Omega(P) = \frac{1}{2\pi i}\int_{-\infty}^\infty dP' \frac{\rho(P')}{P-P'}
\end{equation}
We can also rewrite this in momentum space, which shows that in the upper (lower) half-plane, $\Omega$ is given by the positive (negative) frequency part of $\rho$. This follows from the Fourier transforms
\begin{equation}
	\mathfrak{h}_{P'}(P) = \frac{1}{2\pi i(P-P')} \implies \hat{\mathfrak{h}}_{P'}(P) = \pm \Theta(\mp P) \kernel_{PP'},\quad \Im P' \gtrless 0,
\end{equation}
where $\kernel$ is the Fourier kernel \eqref{eq:modS}, and $\Theta$ the Heaviside step function.

We can also perform the Fourier transform directly on analytic representations. If $\Omega$ is an analytic representation of $\rho$, by a slight abuse of notation we denote an analytic representation of $\hat{\rho}$ by $\hat{\Omega}$:
\begin{equation}\label{eq:OmegaHat}
\hat{\Omega}(P) = \int_{\Gamma_\pm} dP' \kernel_{P P'}\Omega(P') \quad \Im P \gtrless 0
\end{equation}
The contours $\Gamma_\pm$ each consist of two pieces, one in the upper half-plane going from $\Re P\to \mp\infty$ to an arbitrary point $P_U$ (in the domain of definition of $\Omega$), and one in the lower half-plane running similarly from an arbitrary point $P=P_L$ to $\Re P\to \mp\infty$. Changing $P_U$ or $P_L$ amounts to adding identical contours to $\Gamma_\pm$, which adds an entire function to $\Omega_{\hat{\rho}}$, leaving $\hat{\rho}$ invariant as required. The splitting of the contour for $\Im P>0$ and $\Im P<0$ is designed to avoid exponential growth in the integral from the Fourier kernel $\kernel$.

To use analytic representations, instead of choosing $\rho$ to be even it may be convenient to choose it instead to have no support for negative $P$. Then, $\Omega$ will be analytic in the whole left half-plane. The contours in the integrals \eqref{eq:OmegaHat} can then be closed, with $\Gamma_+$ becoming empty and $\Gamma_-$ going from $\Re P\to \infty$ in the upper half-plane to $\Re P\to \infty$ in the lower half-plane, looping round all singularities. For application to the S-transform this must then be combined with the conjugate, since we there assumed that $\rho$ was even.

We note that the analytic function $C(\Delta,J)$ appearing in \cite{1703.00278}, which encodes the OPE coefficients of a correlation function in its poles, is an analytic representation of the spectral density in the sense described here. Perhaps existing mathematical results can be helpful for uncovering the physical consequences of analyticity in spin.

\subsection{A distributional binomial theorem}

In section \ref{eq:sumspectrum}, we used a binomial theorem applied to distributions:
\begin{equation}
	\sum_{n=0}^\infty \binom{\tfrac{1}{2}}{n} \eta^{2n} \pv \frac{1}{P^{2n-2}} \stackrel{?}{=} |P|\sqrt{P^2+\eta^2}
\end{equation}
In this section, we derive this sum, and clarify the meaning of the distribution on the right hand side.

This serves as an example of the power of the analytic representations of the previous subsection: we will perform the sum over the analytic representation
\begin{equation}
	-\tfrac{1}{2} P^{-2k} \sgn{\Im P}
\end{equation}
of $\pv \frac{1}{P^{2n-2}}$. In particular, this will automatically take into account the regularisation implied by the principal value symbol.

Now we can perform the sum on this analytic function, and it converges (uniformly on compacta) for $|P|>\eta$, to
\begin{align*}
	\Omega_\eta &= \sum_{n=0}^\infty \binom{\tfrac{1}{2}}{n} \eta^{2n} \left[-\tfrac{1}{2}\frac{1}{P^{2n-2}}\sgn{\Im P}\right] \\
	&= -\tfrac{1}{2}P^2\sqrt{1+\frac{\eta^2}{P^2}} \,\sgn{\Im P} \\
	&= \frac{1}{2i}P\sqrt{-(\eta^2+P^2)},
\end{align*}
where we take the principal branch of the square root, with cut along the negative real axis.

Note that the terms in the sum are conventional (tempered) distributions, but the series does not converge in that space: this is indicated by the fact that the sum of analytic representations is not analytic everywhere in upper and lower half-planes.  Nonetheless, because our test functions are entire and decay rapidly, convergence in the given region is enough to imply that the sum of distributions converges to $\rho_\eta$, with
\begin{equation}
	\langle\rho_\eta,\psi\rangle := \frac{1}{2i} \int_\Gamma dP\; \psi(P) P\sqrt{-(\eta^2+P^2)}.
\end{equation}
The contour $\Gamma$ runs from left to right in the lower half-plane and right to left in the upper half-plane as above, here in particular staying in the analytic region $|\Im P|>\eta$ to avoid the branch cut. 

We can deform the contour to run along the cuts, so it just picks up the discontinuity. For example, on the positive real axis, the integrand is $-i\psi(P)P\sqrt{P^2+\eta^2}$ just above the cut, and the negative of this just below it, so this contributes $\int_0^\infty dP\; \psi(P) P\sqrt{\eta^2+P^2}$. But there is also a contribution from jumps across branch cuts on the imaginary axis between $\pm i \eta$:
\begin{equation}
	\langle\rho_\eta,\psi\rangle =  \int_{-\infty}^\infty dP\; \psi(P) |P|\sqrt{\eta^2+P^2}+ \int_{-\eta}^\eta dy\, \psi(i y) |y|\sqrt{\eta^2-y^2}
\end{equation}
After translating from $P$ variables to $h$, this gives a density  of states going like $\sqrt{h-\frac{c-1}{24}+\eta^2}$, starting at the zero of the square root, not just at $h=\frac{c-1}{24}$.

\subsection{Asymptotics of distributions}

An important part of our analysis is a characterisation of the asymptotic behaviour of distributions at large $|P|$, corresponding to large spin. Since we are interested in distributions supported on discrete sets of points, it is not immediately clear how to make sense of this, in particular because the usual definition of an asymptotic series fails for such distributions. For us, we want to know what conditions on a distribution suffice to determine the most singular terms in its Fourier transform.

We illustrate various possible approaches to this problem with a simple example, taking a distribution similar to the density of states on double twist Regge trajectories:
\begin{equation}
	\rho(P) = \sum_{\substack{\ell=0\\ \text{even}}}^\infty \left[\delta\left(P-\sqrt{\ell}\right) + \delta\left(P+\sqrt{\ell} \right) \right]
\end{equation}

One approach is to integrate the distribution several times, until it becomes a smooth enough function to use classical results on Fourier transforms. The strategy is to integrate $\rho$ repeatedly, constructing $\rho_k$ with $k$th derivative $\rho$. If $\rho$ is sufficiently well-behaved, we will find that, for large enough $k$, $\rho_k$ will be a simple function $p_k$ ($\sgn(P)$ times a polynomial in cases of interest here) plus a remainder $r_k$ which decays sufficiently quickly at infinity. In particular, if $r_k$ is absolutely integral ($r_k\in L^1(\RR)$, $\int_{-\infty}^\infty dP \, |r_k(P)| < \infty$), then it has a bounded and continuous Fourier transform. We then have the asymptotic estimate
\begin{equation}
\hat{\rho}(P) - (-4\pi i P)^k \hat{\rho}_k(P) = O(P^k) \qquad \text{as }P\to 0,
\end{equation}
and the difference of distributions on the left is in fact a continuous function.

In our example, we can integrate once to get the total number of states below a given $P$,
\begin{equation}\label{eq:rho1App}
	\rho_1(P) = \int \rho(P) = \sgn (P) \left\lfloor \frac{P^2}{2}+1 \right\rfloor,
\end{equation}
but subtracting simple functions like $\sgn(P)(\frac{P^2}{2}+c)$ does not result in a decaying function, since there are oscillations of a constant size as $P\to\infty$. Integrating again improves things, with oscillations decaying like $P^{-1}$ at large $P$, but still $r_2\notin L^1$. In fact, we must integrate three times, to get
\begin{equation}
	\rho_3(P) = \sgn(P)\left[ \frac{P^4}{4!} + \frac{P^2}{4} -\frac{1}{12} \right] + r_3(P),
\end{equation}
where (with the appropriate choice of constants of integration) $P^2|r_3(P^2)|$ is bounded, so $r_3\in L^1(\RR)$. This gives the result
\begin{equation}
	\hat{\rho}(P) = -\pv \frac{1}{\sqrt{2}(2\pi P)^2} + \sqrt{2} -\frac{2}{3}\sqrt{2} (2\pi P)^2 + O(P^3).
\end{equation}
We could integrate yet more times to find higher orders in the expansion, but the calculations quickly become unwieldy.

The size of the fluctuations in the remainders of integrated densities $r_k$ are directly related to the typical spacing between states as $P\to\infty$. A very uneven spectrum, with clumps of many nearby or degenerate states interspersed by large gaps, will tend to require larger $k$ for the arguments outlined here to succeed. It would be interesting to pursue this more systematically.

In some circumstances, analytic representations can be a powerful tool for computing asymptotic expansions of the discretely supported distributions of interest. For our example distribution, we find an analytic representation by summing the poles representing delta-functions, subtracting some entire analytic pieces for convergence and convenience:
\begin{equation}
	\Omega(P) =\sum_{\substack{\ell=0\\ \text{even}}}^\infty  \frac{1}{2i\pi}\left[ \frac{2P}{P^2-\ell}+\frac{2P}{\ell+2}\right] -\frac{\gamma}{2\pi i} P = \frac{1}{2\pi i}P \psi\left(-\tfrac{1}{2}P^2\right)	
\end{equation}
Here, $\psi$ is the digamma function, which has the asymptotic expansion
\begin{equation}
	\psi(z)\sim \log z-\frac{1}{2z} -\sum_{n=1}^\infty \frac{B_{2n}}{2n z^{2n}}
\end{equation}
valid for $|z|\to\infty$ anywhere away from the negative real $z$ axis, where the poles of $\psi$, which condense into the $\log$ branch cut, are located. This gives an asymptotic expansion for our analytic representation of $\rho$:
\begin{equation}
	\Omega(P) \sim \frac{P}{2\pi i} \psi\left(-\tfrac{1}{2}P^2\right) \sim \frac{1}{2\pi i} \left[ P \log\left(-\tfrac{1}{2}P^2\right)+\frac{1}{P}-\sum_{n=1}^\infty \frac{2^{2n}B_{2n}}{2n P^{4n-1}}\right]
\end{equation}
This is valid in a limit $|P|\to\infty$, as long as $\Im P \to \pm\infty$ (it fails taking $|P|\to\infty$ along lines parallel to the real axis). Na\"ively interpreting as a distribution term-by-term, we can write a formal asymptotic expansion of $\rho$:
\begin{equation}
	\rho(P) \sim |P| + \delta(P) + \cdots
\end{equation}
The first term is very reasonable, giving the density of states at large $P$. The $\delta$-function also has a nice interpretation, giving the average discrepancy between the total number of states below a given $P$ \eqref{eq:rho1App}, and the number $P^2/2$ deduced from the leading term.

We now take the Fourier transform. Computationally, it is simplest to interpreting the terms of the expansion as derivatives of delta functions and transform term by term. However, this hides that we are rigorously computing a small $P$ expansion of the analytic representation $\hat{\Omega}$ of $\hat{\rho}$, using \eqref{eq:OmegaHat}. We get the expansion
\begin{equation}
	\hat{\rho}(P) = -\frac{1}{\sqrt{2}(2\pi P)^2} + \sqrt{2}+\sqrt{2}\sum_{n=1}^\infty \frac{2^{2n}B_{2n}}{2n (4n-2)!} (4\pi P)^{4n-2}.
\end{equation}
Despite the sum in intermediate steps being asymptotic, this in fact has infinite radius of convergence, and really equals the Fourier transform. We can check this directly by using the Gaussian test functions $\chi_P$ in this case (which amounts to constructing the partition functions for $\rho$, $\hat{\rho}$ and checking they are related by $\tau\to - 1/\tau$).

\section{Counting multi-trace states}\label{app:counting}

In this appendix, we enumerate the operators of mean field theory, which form a Fock space built from the primary $\phi$ and its global descendants $\partial^k\bar{\partial}^{\bar{k}}\phi$. These results are applied in the text to count multi-twist operators.

\subsection{Bosons}

Begin with the Bose-Einstein partition function:
\begin{equation}
	\pf_\text{Bose} = \prod_{k,\bar{k}=0}^\infty \frac{1}{1-q^{h_\phi+k}q^{\bar{h}_\phi+\bar{k}}}.
\end{equation}
To massage this into a more useful form, we take the log to write the product as a sum, expand the terms using the Taylor series for $\log$, and perform the sum over $k,\bar{k}$:
\begin{align*}
\log\pf_\text{Bose} &= 	-\sum_{k,\bar{k}=0}^\infty \log\left(1-q^{h_\phi+k}\bar{q}^{\bar{h}_\phi+\bar{k}}\right) \\
&= 	\sum_{k,\bar{k}=0}^\infty\sum_{n=1}^\infty \frac{1}{n} q^{(h_\phi+k)n}\bar{q}^{(\bar{h}_\phi+\bar{k})n} \\
&= \sum_{n=1}^\infty \frac{1}{n}\frac{q^{h_\phi n}\bar{q}^{\bar{h}_\phi n}}{(1-q^n)(1-\bar{q}^n)}
\end{align*}
This has a nice bulk worldline QFT interpretation: we're exponentiating all connected diagrams, which for free particles  just consist of closed loops. The index $n$ counts the number of times a loop winds round the thermal circle (the $n=0$ term is cancelled by renormalisation of the cosmological constant). The $1/n$ is a symmetry factor, multiplying the single-particle partition function with $\beta\rightarrow n\beta$.

Now, exponentiate and Taylor expand each term:
\begin{equation}
	\pf_\text{Bose} =  \prod_{n=1}^\infty \sum_{k=0}^\infty \frac{1}{k!} \left(\frac{q^{h_\phi n}\bar{q}^{\bar{h}_\phi n}}{n(1-q^n)(1-\bar{q}^n)}\right)^k
\end{equation}
The contribution from a given particle number $p$ will be a sum of terms that are labelled by partitions of $p$: for each $n$, we pick the $k_n$th term in the sum over $k$, where $k_n$ is the number of times $n$ appears in the partition.
\begin{equation}
	\pf_p = \sum_{\substack{\text{Partitions}\\ \text{of $p$}}} \prod_{n=1}^\infty \frac{1}{k_n! \left(n(1-q^n)(1-\bar{q}^n)\right)^{k_n}},
\end{equation}
where
\begin{equation}
	\pf_\text{Bose} = \sum_{p=0}^\infty \pf_p q^{p h_\phi}\bar{q}^{p \bar{h}_\phi}.
\end{equation}
For small particle numbers, we have the following:
\begin{align*}
	&\quad\pf_1 =\ydiagram{1} = \frac{1}{(1-q)(1-\bar{q})},\quad \pf_2 = \ydiagram{1,1}+\ydiagram{2}=\frac{1}{2(1-q)^2(1-\bar{q})^2} + \frac{1}{2(1-q^2)(1-\bar{q}^2)} \\
	\pf_3 &= \ydiagram{1,1,1} + \ydiagram{2,1} + \ydiagram{3} = \frac{1}{6(1-q)^3(1-\bar{q})^3} + \frac{1}{2(1-q)(1-\bar{q})(1-q)^2(1-\bar{q})^2} + \frac{1}{3(1-q^3)(1-\bar{q}^3)}
\end{align*}

To count primaries only, multiply by $(1-q)(1-\bar{q})$ to subtract descendants. For example, for two-particle states we have
\begin{equation}
	(1-q)(1-\bar{q})\pf_2 = \frac{1}{2}\left[\frac{1}{(1-q)(1-\bar{q})}+\frac{1}{(1+q)(1+\bar{q})}\right].
\end{equation}
The first term gives an operator of every spin for each twist; the second cancels the odd spins to give a single operator at each even spin.

We observe experimentally that the leading Regge trajectory for $p$ particles is given by the simple generating function
\begin{equation}
	\left. \pf_p \right|_{q=0} = \prod_{k=1}^p \frac{1}{1-\bar{q}^k},
\end{equation}
which means that number of primaries in the leading $p$-particle Regge trajectory at each spin $\ell$ is given by the number of partitions of $\ell$ into $2,3,4,\cdots,p$.

\subsection{Counting at large spin}

The growth of degeneracies at large spin is determined by the $\bar{q}\to 1$ limit, which for a given Young tableau is controlled by the height (the number of elements in the partition). The leading order is therefore all given by the partition of $p$ into lots of ones, $\frac{1}{p!(1-q)^p(1-\bar{q})^p}$. The degeneracies of primaries from this contribution is then given by the binomial formula,
\begin{equation}
	\frac{1}{p!}\frac{(p+m-2)!}{m!(p-2)!}\frac{(p+\bar{m}-2)!}{\bar{m}!(p-2)!},
\end{equation}
which is the coefficient of $q^m \bar{q}^{\bar{m}}$ in $\frac{1}{p!(1-q)^{p-1}(1-\bar{q})^{p-1}}$. Taking the large spin asymptotics at fixed $m$ with $\bar{m}=m+\ell$, we get degeneracies
\begin{equation}
	 \frac{(p+m-2)!}{m!(p-2)!} \frac{\ell^{p-2}}{(p-2)! p!} \:.
\end{equation}
While this is not accurate for each individual spin, it is correct for the total integrated number of operators,
\begin{equation}
	 N_{p,m}(\ell)\sim \frac{(p+m-2)!}{m!(p-2)!} \frac{\ell^{p-1}}{(p-1)! p!} \:.
\end{equation}

To prove this, we can apply a Tauberian theorem to the coefficients of $q^m$ in the $p$-particle partition function $(1-q)(1-\bar{q})\pf_p(q,\bar{q})$. The $\bar{q}\to 1$ asymptotics are determined by the term in the sum over partitions described above. The Hardy-Littlewood Tauberian theorem\footnote{\href{https://www.encyclopediaofmath.org/index.php/Tauberian_theorems}{\texttt{https://www.encyclopediaofmath.org/index.php/Tauberian\_theorems}}} then implies that the sum of coefficients of $\bar{q}^{\bar{m}}$ for $\bar{m}\leq \ell$ obeys the given asymptotic formula.

\subsection{Multiple species}

For multiple species of particles, just take the product of Bose partition functions for each. For the asymptotic large spin piece of the partition function with particle numbers $p_i$, this gives
\begin{equation}
	\pf_{\{p_i\}} \sim \prod_i \frac{1}{p_i!}\frac{1}{(1-q)^{p_i}(1-\bar{q})^{p_i}} = \binom{p}{p_1,\cdots,p_N} \frac{1}{p!}\frac{1}{(1-q)^{p}(1-\bar{q})^{p}}
\end{equation}
where $p=\sum p_i$, which is just the multinomial coefficient times the single-species, $p$ particle answer. This simple result only holds to leading order at large spin.

\subsection{Fermions}

We here look at Fermi statistics. The Fock space of `light' fermions we're considering will always be in the NS (antiperiodic) sector. But we can still put periodic or antiperiodic boundary conditions in the thermal cycle; the former means we include a $(-1)^F$ insertion. We'll therefore look at two partition functions $\pf^\pm =\Tr((\pm)^F e^{-\beta H})$:
\begin{equation}
	\pf^\pm_\text{Fermi} = \prod_{k,\bar{k}=0}^\infty (1\pm q^{h_\phi+k}q^{\bar{h}_\phi+\bar{k}})
\end{equation}
Following the same steps, we get
\begin{equation}
	\log \pf^\pm_\text{Fermi} = -\sum_{j=1}^\infty \frac{(\mp)^{j}}{j}\frac{q^{h_\phi j}\bar{q}^{\bar{h}_\phi j}}{(1-q^j)(1-\bar{q}^j)},
\end{equation}
and so
\begin{equation}
	\pf^\pm_\text{Fermi} =  \prod_{j=1}^\infty \sum_{k=0}^\infty \frac{1}{k!} \left(-\frac{(\mp)^{j}}{j}\frac{q^{h_\phi j}\bar{q}^{\bar{h}_\phi j}}{(1-q^j)(1-\bar{q}^j)}\right)^k.
\end{equation}
Now the $p$-particle sector will involve the same sum over partitions, but with extra signs.
Writing
\begin{equation}
	\pf^\pm_\text{Fermi} = \sum_{p=0}^\infty(\pm)^p \pf^\text{F}_p\, q^{p h_\phi}\bar{q}^{p \bar{h}_\phi},
\end{equation}
we have
\begin{equation}
	\pf^\text{F}_p = (-1)^p \!\!\! \sum_\text{Partitions of $p$} \prod_j \frac{(-1)^{k_j}}{k_j! \left(j(1-q^j)(1-\bar{q}^j)\right)^{k_j}},
\end{equation}
Compared to the Bose case, we have extra signs $(-1)^{p+\sum k_j}$ weighting different partitions. For example, for $p=2$, we get
\begin{equation}
	\pf_2 = \ydiagram{1,1}-\ydiagram{2}=\frac{1}{2(1-q)^2(1-\bar{q})^2} - \frac{1}{2(1-q^2)(1-\bar{q}^2)},
\end{equation}
which projects onto even spin primaries (since the (absent) $q^0\bar{q}^0$ term would correspond to an odd spin).

The leading order piece at large spin is given by the same result as for bosons: the partition into $p$ ones has $\sum k_j=p$, so that contribution to $\pf^\text{F}_p$ comes without a sign, as it must for positive degeneracies.

\section{The covariant phase space formalism\label{app:Wald}}

\subsection{The formalism}

We consider a general theory in $d+1$ dimensions with dynamical fields $\phi$ (including the metric), and diffeomorphism invariant Lagrangian, which we express as a $(d+1)$-form $\mathbf{L}$. Under a general variation, we have
\begin{equation}
	\delta \mathbf{L} = \mathbf{E}(\delta\phi) + d\mathbf{\Theta}(\delta\phi).
\end{equation}
The two terms are the equations of motion for the background $\mathbf{E}$, expressed as a linear function of the field variations, and the symplectic potential $\mathbf{\Theta}$, a $d$-form depending linearly on the field variations and perhaps their derivatives.

The diffeomorphism invariance of the theory means that, for any infinitesimal diffeomorphism labelled by a vector field $\xi$, we have
\begin{equation}
\delta_\xi \phi	= \mathcal{L}_\xi \phi \implies \delta_\xi \mathbf{L} = \mathcal{L}_\xi \mathbf{L}.
\end{equation}
For any $\xi$, there is an associated Noether current\footnote{We use $\iota$ to denote the interior product, $(\iota_X \alpha)_{a_1 a_2\cdots}= X^b\alpha_{ba_1a_2\cdots}$.} $d$-form $\mathbf{J}_\xi$,
\begin{equation}
	\mathbf{J}_\xi = \mathbf{\Theta}(\mathcal{L}_\xi\phi)-\iota_\xi\mathbf{L},
\end{equation}
which satisfies the conservation equation
\begin{equation}
	d\mathbf{J}_\xi = -\mathbf{E} (\mathcal{L}_\xi\phi) \quad (=0 \quad \text{on-shell)}.
\end{equation}
From this, there is a $(d-1)$-form $\mathbf{Q}_\xi$, constructed locally from the fields and $\xi$, such that
\begin{equation*}
	d\mathbf{Q}_\xi = \mathbf{J}_\xi \quad (\text{on-shell}).
\end{equation*}

We now wish to construct generators of symmetries and conserved quantities from $\mathbf{Q}_\xi$. To find the correct object, we consider $\mathbf{J}_\xi$ under field variations. A short calculation gives
\begin{equation*}
	\delta\mathbf{J}_\xi = d(\iota_\xi \mathbf{\Theta}(\delta\phi))- \iota_\xi \mathbf{E}(\delta\phi) + \delta\left[\mathbf{\Theta}(\mathcal{L}_\xi \phi)\right]-\mathcal{L}_\xi\left[\mathbf{\Theta}(\delta\phi)\right],
\end{equation*}
and the last two terms can be written in terms of the symplectic current
\begin{equation}
\omega(\delta_1\phi,\delta_2\phi) = 	\delta_1\mathbf{\Theta}(\delta_2 \phi) - \delta_2\mathbf{\Theta}(\delta_1 \phi),
\end{equation}
as $\omega(\delta\phi,\mathcal{L}_\xi\phi)$ (a pair of terms with $\mathcal{L}_\xi\delta\phi$ cancel). The symplectic form is the integral of $\omega$ on a Cauchy surface. This motivates the construction of the ($d-1$)-form
\begin{equation}
 \delta\mathbf{H}_\xi	=\delta\mathbf{Q}_\xi - \iota_\xi \mathbf{\Theta}(\delta\phi),
\end{equation}
which is the variation of the Hamiltonian density generating translation by $\xi$: it is related, through the symplectic form, with $\delta\phi$ (regarded as a vector field on phase space) by Hamilton's equations:
\begin{equation*}
	 d\delta\mathbf{H}_\xi = \omega(\delta\phi,\mathcal{L}_\xi\phi) \quad (\text{on-shell})
\end{equation*}
In particular, this vanishes if $\xi$ is a symmetry of the background configuration, $\mathcal{L}_\xi\phi=0$.

 The integral of $\delta\mathbf{H}_\xi$ on a spacelike $(d-1)$-dimensional boundary at infinity defines the variation of conserved charges associated with $\xi$, in particular the energy and angular momentum for $\xi=\partial_t,\partial_\phi$ respectively. Since this is a closed form, it can be evaluated on any homologous $(d-1)$-dimensional submanifold; in particular, comparing the evaluation at infinity with a horizon gives the first law of black hole thermodynamics.

\subsection{Einstein-Hilbert}

We now follow this construction for pure Einstein-Hilbert gravity (in $d+1$ spacetime dimensions), taking
\begin{equation}
\mathbf{L} = (R-2\Lambda)\epsilon,	
\end{equation}
where $\epsilon$ is the volume form. We have excluded the normalisation factor $\frac{1}{16\pi G_N}$ here to reduce clutter; it is restored in the main text.

Computing the variation with respect to the metric, $\delta g_{ab} = h_{ab}$ we find
\begin{align}
\mathbf{E}(h) &= -E^{ab}h_{ab} \epsilon,\quad E^{ab}=R^{ab}-\tfrac{1}{2}Rg^{ab}+\Lambda g^{ab} \\
\mathbf{\Theta} &= \iota_X\epsilon, \quad X^a = (g^{ac}g^{bd}-g^{ab}g^{cd})\nabla_b h_{cd}.
\end{align}
From this, we construct the Noether current for a vector field $\xi$,\footnote{Notation: $\star$ is the Hodge dual $(\star\alpha)_{a_{p+1}\cdots a_{d+1}} = \tfrac{1}{p!}\alpha^{a_1\cdots a_p}\epsilon_{a_1\cdots a_{d+1}}$; the flat $\flat$ denotes the one-form dual to a vector, $(X^\flat)_a=g_{ab}X^b$; the codifferential $\delta$ acts as $(-1)^p\star^{-1}d\star$ on $p$-forms, and in coordinates gives $(\delta \alpha)_{a_1\cdots a_{p-1}}=-\nabla^b\alpha_{ba_1\cdots a_{p-1}}$.}
\begin{equation}
\mathbf{J}_\xi = 2\star E\cdot\xi - d\star d\xi^\flat	, \qquad (E\cdot \xi	 = E_{ab}\xi^b dx^a),
\end{equation}
and read off the Noether charge
\begin{equation}
	\mathbf{Q}_\xi = -\star d\xi^\flat,
\end{equation}
which satisfies
\begin{equation}
	d\mathbf{Q}_\xi = \mathbf{J}_\xi-2\star E\cdot\xi.
\end{equation}
We now consider the Hamiltonian associated with a Killing field $\xi$,
\begin{equation}
	\delta\mathbf{H}_\xi = -\delta(\star d\xi^\flat) - \iota_\xi\iota_X\epsilon.
\end{equation}
In a background satisfying the Einstein equations, but with an arbitrary (off-shell) first order variation with linearised Einstein tensor $E_{ab}$, we have the following:
\begin{equation}
d\delta\mathbf{H}_\xi= -2\star E\cdot \xi
\end{equation}
Restoring the factor of $16\pi G_N$, this gives the conservation equation \eqref{eq:conservation}.

\bibliographystyle{JHEP}
\bibliography{NearExtremalBootstrap}

\end{document}